\documentclass[10pt,twocolumn,a4paper]{IEEEtran}

\usepackage[mathscr]{euscript}
\let\euscr\mathscr \let\mathscr\relax
\usepackage[scr]{rsfso}
\usepackage[active]{srcltx} 
\usepackage[cmex10]{amsmath}
\usepackage{amsfonts}
\usepackage[final]{graphics}
\usepackage[final]{graphicx}
\usepackage{amsbsy}
\usepackage{amsbsy}
\usepackage{amssymb}
\usepackage{url}
\usepackage{enumerate}
\usepackage{cite}
\usepackage{psfrag}
\usepackage{color}
\usepackage{euscript}
\usepackage[utf8]{inputenc}
\usepackage{xcolor}
\usepackage{tabularx}
\usepackage{algorithm}
\usepackage{algpseudocode} 

\newcommand{\bm}[1]{{\mathbf{#1}}}
\newcommand{\Es}{{\mathbb{E}}}          

\newcommand{\Cset}{\mathbb{C}}

\newcommand{\eqdef}{\triangleq}

\newcommand{\herm}{\text{H}}

\newcommand{\rate}{\EuScript{R}}

\def\bdm#1\edm{\begin{displaymath}#1\end{displaymath}}
\def\be#1\ee{\begin{equation}#1\end{equation}}
\def\barr#1\earr{\begin{align}#1\end{align}}

\newcommand{\IeeeTIT}{{\em IEEE Trans.\ Inf. Theory\/}}
\newcommand{\IeeeTSP}{{\em IEEE Trans.\ Signal Process.\/}}
\newcommand{\IeeeTCOMM}{{\em IEEE Trans.\ Commun.\/}}

\newcommand{\IeeeWCOMMLETT}{{\em IEEE Wireless Commun.\ Lett.\/}}
\newcommand{\IeeeTWC}{{\em IEEE Trans.\ Wireless Commun.\/}}
\newcommand{\IeeeJSAC}{{\em IEEE J.\ Select.\ Areas Commun.\/}}
\newcommand{\IeeeTVT}{{\em IEEE Trans.\ Veh. Technol.\/}}

\usepackage{tikz}

\newcommand\copyrighttext{%
  \footnotesize \the\year{} IEEE. Personal use of this material is permitted. Permission from IEEE must be obtained for all other uses, including reprinting/republishing this material for advertising or promotional purposes, collecting new collected works for resale or redistribution to servers or lists, or reuse of any copyrighted component of this work in other works.}
\newcommand\copylefttext{%
  \footnotesize This work is licensed under a Creative Commons Attribution 4.0 License. For more information, see https://creativecommons.org/licenses/by/4.0.}

\newcommand\copyrightnotice{%
\begin{tikzpicture}[remember picture,overlay]
\node[anchor=south,yshift=10pt] at (current page.south) {%
\begin{minipage}{\textwidth}
\center \copylefttext
\end{minipage}};
\end{tikzpicture}%
}

\newcommand\acceptedtext{%
\footnotesize This article has been accepted for publication in IEEE Transactions on Communications. This is the author's version which has not been fully edited and
content may change prior to final publication. Citation information: DOI 10.1109/TCOMM.2026.3686695}
  
\newcommand\acceptednotice{%
\begin{tikzpicture}[remember picture,overlay]
\node[anchor=north,yshift=0pt,xshift=0pt] at (current page.north) {%
\begin{minipage}{\textwidth}
\center \acceptedtext
\end{minipage}};
\end{tikzpicture}%
}

\begin{document}

\title{Randomized Space-Time Stacked Intelligent Metasurfaces for 
Massive Multiuser \\ Downlink Connectivity}

\author{Donatella~Darsena,~\IEEEmembership{Senior Member,~IEEE}, Ivan~Iudice,~\IEEEmembership{Senior Member,~IEEE}, 
Vincenzo~Galdi,~\IEEEmembership{Fellow,~IEEE}, \\ and Francesco~Verde,~\IEEEmembership{Senior Member,~IEEE} 
\thanks{
Manuscript received October 24, 2025; 
revised March 9, 2026;
accepted April 12, 2026.
The associate editor coordinating the review of this paper and
approving  it for publication  was Dr.~Chao-Kai Wen.
(\em Corresponding author: Francesco Verde)
}
\thanks{
D.~Darsena is with the Department of Electrical Engineering and Information Technology,  University Federico II, Naples I-80125,
Italy (e-mail: darsena@unina.it).
I.~Iudice is with the Reliability \& Security Department, Italian Aerospace Research Centre (CIRA),
Capua I-81043, Italy (e-mail: i.iudice@cira.it).
V.~Galdi is with the Department of Engineering, University of Sannio,  
Benevento I-82100, Italy (e-mail: vgaldi@unisannio.it).
F.~Verde is with the Department of Engineering, 
University of Campania Luigi Vanvitelli, Aversa I-81031, Italy
(e-mail: francesco.verde@unicampania.it).
}
\thanks{
The work of D.~Darsena, I.~Iudice, and V.~Galdi was partially supported by 
the European Union-Next Generation EU under the Italian
National Recovery and Resilience Plan (NRRP), Mission 4,
Component 2, Investment 1.3, CUP E63C22002040007, partnership
on ``Telecommunications of the Future" (PE00000001
- program ``RESTART").
}}
\markboth{IEEE Transactions on Communications, Vol.~xx,
No.~yy,~zz~2026}{Darsena\MakeLowercase{\textit{et al.}}:
Randomized Space-Time Stacked Intelligent Metasurfaces for 
Massive Multiuser Downlink Connectivity}

\IEEEpubid{0000--0000/00\$00.00~\copyright~2026 IEEE}

\maketitle
\acceptednotice
\copyrightnotice

\begin{abstract}

Stacked intelligent metasurfaces (SIMs) represent a key enabler for next-generation wireless networks,  offering beamforming gains 
while significantly reducing the number of radio-frequency chains.
In conventional space-only (S-only) SIM architectures, the rate of reconfigurability of the SIM 
is equal to the inverse of the channel coherence time.
This paper investigates a novel beamforming strategy for massive downlink connectivity using a randomized 
space-time (ST) SIM. In addition to conventional S-only metasurface layers, 
the proposed design integrates an ST metasurface 
layer at the input stage of the SIM that 
introduces random variations over each channel coherence interval.
These artificial time variations enable opportunistic user scheduling and exploitation of multiuser diversity under slow channel dynamics. 
To mitigate the prohibitive 
overhead associated with full channel state information at the transmitter (CSIT), we propose a partial-CSIT-based beamforming 
scheme that leverages randomized steering vectors and limited user-side feedback based on signal quality measurements.
Numerical results  demonstrate that the proposed ST-SIM architecture achieves satisfactory sum-rate performance 
while significantly reducing CSIT acquisition and feedback overhead, thereby enabling scalable 
downlink connectivity in dense networks.

\end{abstract}


\section{Introduction}

\IEEEPARstart{T}{he} ever-increasing demand for ultra-reliable, high-capacity wireless services is driving the development of sixth-generation (6G) networks, 
which aim to enable massive connectivity, low latency, and unprecedented spectral and energy efficiency \cite{ITU.2023,Kalor.2024}. Conventional fully digital beamforming architectures, 
though effective in providing high spatial resolution, face critical scalability issues in dense network deployments. Specifically, the need for a large number 
of radio-frequency (RF) chains and high-resolution digital-to-analog
and analog-to-digital converters
leads to excessive hardware complexity, energy consumption, and cost. These limitations have motivated 
the development of alternative beamforming strategies that can deliver comparable performance while significantly reducing RF hardware requirements.

{\em Stacked intelligent metasurfaces (SIMs)} have emerged as a promising technology to address these challenges \cite{Hanzo}.
A SIM consists of multiple cascaded programmable metasurface layers that can shape the electromagnetic (EM) wavefront directly 
in the propagation domain, effectively implementing analog signal transformations without requiring additional RF chains or digital hardware. 
This concept builds upon diffractive deep neural networks ($\text{D}^2$NN) \cite{Lin.2018,Liu.2022}, where wavefront transformations are realized by carefully engineering 
the transmission coefficients of metasurface layers to achieve a target mapping between input and output fields. 
In wireless communications, SIMs can replace or complement digital beamforming modu\-les, thereby reducing hardware complexity, 
lowering power consumption, and 
enabling real-time wave-domain beamforming.
Compared to reconfigurable intelligent surfaces (RISs) \cite{Basar.2024}, which are typically placed in the environment to control wireless propagation, 
SIM modules are deployed at the transmitter or receiver side and act as active or passive analog beamformers. Such a stacked architecture allows 
for more flexible and efficient transformations, as metasurfaces are part of the transceiver chain.

\IEEEpubidadjcol

\subsection{Related works}

The potential of SIM technology for wireless communications has been explored in several recent studies. 
Some works focused on point-to-point scenarios, analyzing the fundamental wave propagation mechanisms through 
stacked metasurfaces and the resulting beamforming capabilities. A free-space path-loss model for SIM-based 
transmitters was proposed in \cite{Hassan.2024}, while \cite{Nerini.2024} introduced a multiport network model 
accounting for mutual coupling effects. 
Further developments included SIM-assisted direction-of-arrival estimation \cite{DiRenzo} and holographic 
multiple-input multiple-output (MIMO) systems \cite{Hanzo},  where SIMs are integrated at both the transmitter 
and receiver to realize parallel subchannel decomposition.
In \cite{Yao.2024}, the authors tackled the problem of channel estimation in SIM-assisted MIMO systems 
by proposing  a low-overhead estimation protocol and subspace-based linear estimators that leverage the SIM spatial 
correlation structure.
A double-SIM-assisted massive MIMO architecture was proposed in \cite{Pap.2025}, which 
integrates a hybrid SIM at the base station and an additional SIM in the intermediate space, 
jointly optimized through a projected gradient ascent method to maximize uplink spectral 
efficiency under imperfect channel state information (CSI).

Other works have addressed multiuser downlink communications with SIMs, focusing on sum-rate maximization through wave-domain beamforming. 
In \cite{DiRenzo-ICC}, alternating optimization was used to jointly optimize transmit power and SIM transmission coefficients, 
while \cite{Liu.2024} employed deep reinforcement learning to address the non-convexity of the beamforming design. 
In \cite{An_ArXiv_2025}, the authors extended the
SIM-based transceiver in \cite{DiRenzo-ICC} by considering meta-atoms that can only be tuned discretely.
To alleviate the high overhead 
associated with instantaneous CSI, \cite{Lin.2024} and \cite{Pap.2024} proposed a design based on statistical CSI, significantly simplifying 
system operation in slowly varying channels. 
In \cite{Li.2025}, the authors proposed a hybrid transceiver architecture for near-field wideband systems assisted by SIM, introducing a layer-by-layer holographic beamforming 
algorithm combined with minimum-mean-square-error digital precoding to maximize spectral efficiency under realistic phase tuning errors. 
All these studies have considered phase-only SIMs, which are nearly passive and easier to implement but suffer 
from uncontrolled propagation losses across multiple layers. A more recent contribution investigated the inclusion of {\em active} amplitude-controlling 
layers to enhance wave manipulation capabilities and mitigate such internal losses \cite{Dar.2025}.

Recently, the integration of SIMs into cell-free massive MIMO architectures has been studied to address scalability and fronthaul bottlenecks. 
A digital-wave beamforming framework was introduced in \cite{Li_2024}, where SIM-equipped access points enable high beamforming gains with 
fewer RF chains. Alternating optimization algorithms were developed in \cite{Park-arXiv_2025} for the joint design of digital and wave-domain beamforming, 
as well as fronthaul compression, showing that sufficiently deep SIMs can approach the performance of fully digital schemes. 
Other works \cite{Hu_2025,Shi_2025-May,Shi_2025-June} demonstrated SIMs' ability to improve 
both uplink and downlink sum-rate performance while substantially reducing 
hardware and fronthaul costs.

Recent studies have also investigated SIMs for enhancing the spectral and energy efficiency of multiuser systems. In particular, the authors in \cite{Shi-2025} analyzed the energy efficiency of SIM-assisted communications by jointly optimizing digital precoding and wave-based beamforming,
with special emphasis on the impact of the number of metasurface layers under realistic power consumption models. 

Hardware nonidealities constitute an important practical aspect in holographic and SIM-based wireless architectures, including phase and amplitude imperfections at metasurface elements as well as impairments in the associated RF chains. The impact of such nonidealities was investigated for SIM-assisted cell-free networks in \cite{Li-2024}, while impairment-aware holographic metasurface beamforming was analyzed for multi-altitude low earth orbit (LEO) satellite systems in \cite{Li-2025}. Moreover, the performance of reconfigurable holographic surfaces in near-field cell-free scenarios under phase-shift errors and RF impairments was theoretically characterized in \cite{Li-2024-hw}. 

Despite the promising capabilities of SIMs, most of the aforementioned studies 
have focused on {\em space-only (S-only)} metasurface structures,
which are reconfigured at a rate equal to the inverse of the channel coherence time $T$, often assuming the availability of full CSI 
at the transmitter (CSIT). While such approaches have demonstrated significant performance gains 
in both single-user and multiuser settings, their reliance on frequent CSIT acquisition and deterministic beamforming optimization 
makes them difficult to scale in dense networks with slowly varying channels. Moreover, current S-only designs 
primarily aim to maximize spectral efficiency through static wave-domain beamforming over each time interval of duration $T$, without fully 
exploiting the temporal dimension to enhance scheduling 
flexibility and multiuser diversity. As a result, there remains a gap between the theoretical potential of SIMs and their practical 
deployment in large-scale wireless environments.

\subsection{Contributions}

Our main contributions are summarized as follows:

\begin{enumerate}

\itemsep =1mm

\item
We introduce a novel {\em randomized space–time (ST)} SIM architecture that incorporates a rapidly time-varying (TV) dimensional adaptation layer,
which is randomly reconfigured at a rate greater than $1/T$, 
and multiple S-only metasurface layers, whose characteristic parameters vary slowly at a rate of $1/T$.
This design enables joint spatial-temporal wavefront control, introducing artificial 
time variations over each channel coherence time interval that enhance multiuser diversity even under slowly varying propagation conditions.

\item 
To reduce the prohibitive overhead of full CSIT acquisition, we propose a {\em partial-CSIT} beamforming strategy that leverages randomized 
steering vectors and low-rate signal quality feedback from users. This approach allows for scalable system operation in 
dense networks while maintaining satisfactory sum-rate performance.

\item 
Through {\em extensive numerical simulations}, we show that the proposed ST-SIM architecture achieves significant 
performance gains compared to conventional S-only SIM designs. In particular, 
it approaches full-CSIT beamforming performance for large user populations while significantly reducing signaling overhead,
hence demonstrating its potential for scalable and efficient massive 
downlink connectivity.

\end{enumerate}

It is worth noting that the proposed ST-SIM should not be confused with the space–time coding metasurfaces introduced in 
\cite{Zhang.2018}, which exploit periodic, phase-quantized time modulation to engineer harmonic spectra and control frequency conversion. In contrast, the ST layer considered in this work introduces randomized phase variations across time slots in order to induce artificial channel fluctuations and enable opportunistic multiuser scheduling.

\subsection{Paper organization}
The remainder of this paper is organized as follows. 
Section~\ref{sec:ST-SIM} introduces the system model and problem formulation for the proposed ST-SIM architecture. 
Section~\ref{sec:rx-signal} outlines the mathematical models of both transmitted and received signals.
Section~\ref{sec:random-ST-SIM} presents the partial-CSI scheduling strategy.
In Section~\ref{sec:synthesis}, the optimization framework for designing the SIM transmission coefficients
is developed relying on the gradient descent algorithm. 
Section~\ref{sec:simul} provides illustrative numerical results and performance comparisons with conventional beamforming schemes. 
Finally, Section~\ref{sec:concl} provides some concluding remarks.

\begin{figure*}[t]
\centering
\includegraphics[width=\linewidth]{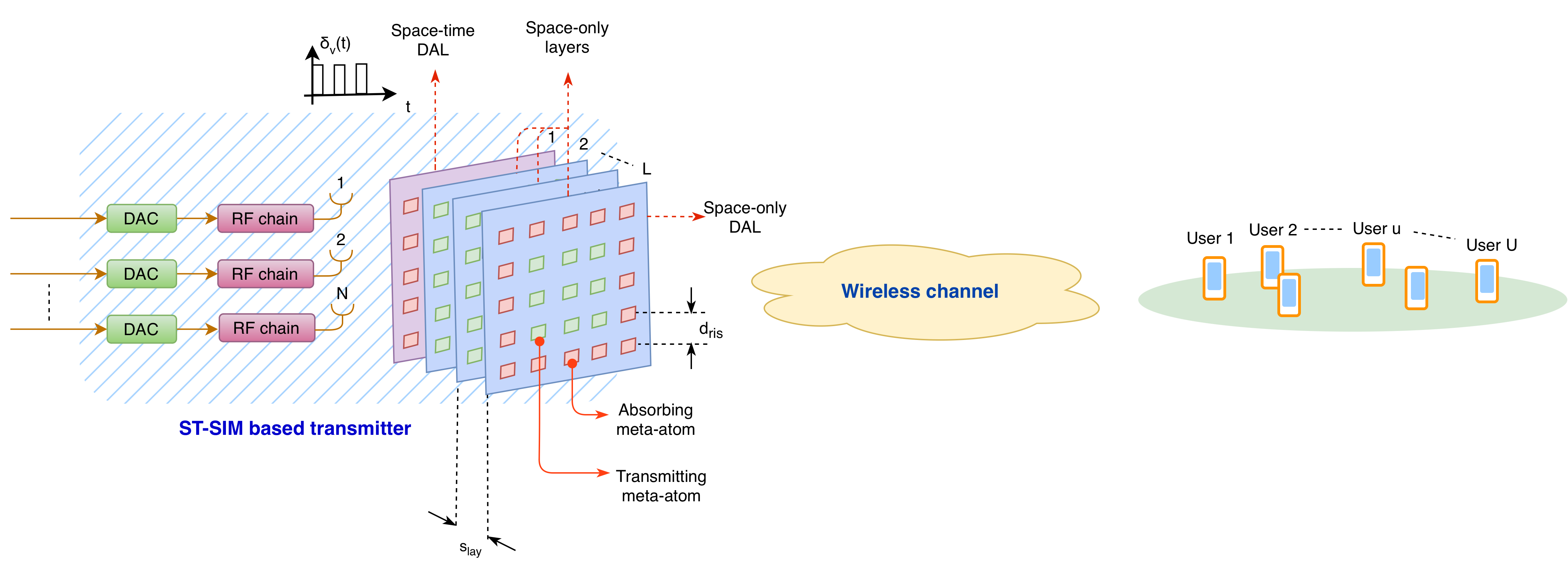}
\caption{ST-SIM-aided multiuser downlink system serving $N$ users out of $U$. 
The ST-SIM consists of $L$ metasurface layers. The first layer acts as a ST-DAL, comprising both absorbing (in red color) 
and transmitting (in green color) meta-atoms. The last $L-1$ layers are S-only ones and consist of transmitting meta-atoms.
S-only layers do not vary over each channel coherence interval of duration $T$, while the ST initial layer is reconfigured at a rate $M/T$.
}
\label{fig:fig_1}
\end{figure*}

\section{Space-time SIM architecture}
\label{sec:ST-SIM}

In this section, we introduce the system model underpinning the proposed ST-SIM architecture.
The baseband modulation spectrum is confined to a 
frequency interval $\mathcal{W}_{\mathrm m}$ of width $B_{\mathrm m}$ (bandwidth) centered at $f=0$, 
with $B_{\mathrm m} \ll f_0$ (narrowband assumption), where $f_0$ denotes the carrier frequency. 
Let $T$ denote the channel coherence time, which is inversely related to the bandwidth
of the Doppler spectrum of the underlying physical channel.
We consider a transmission interval $[0,T)$,  which is partitioned into $M$ time slots 
of duration $T_{\text{s}}$, i.e., $T = M \, T_{\text{s}}$.
We utilize a uniform planar array (UPA) to radiate the
information streams, which consists of 
$N \eqdef N_{x} \times N_{y}$ transmit antennas 
arranged in a rectangular grid with $N_{x}$ and $N_{y}$ elements along the 
$x$ and $y$ axes, respectively, and inter-element spacing
$d_{\text{upa}}$. 
As depicted in Fig.~\ref{fig:fig_1}, 
the base station (BS) is composed of the UPA and the ST-SIM.\footnote{In Fig.~\ref{fig:fig_1}, the UPA is depicted as a linear array of $N$ antennas with inter-element spacing $d_{\text{upa}}$.}
Such a ST-SIM-based transmitter operates on a slot-by-slot basis.

The ST-SIM comprises \( L \) planar layers, 
uniformly spaced by a distance \( s_\text{lay} \) (see Fig.~\ref{fig:fig_1}). 
Specifically, it consists of two functional blocks with {\em different rates of reconfigurability}:
(i) a ST block (in light-purple color in Fig.~\ref{fig:fig_1}), implemented by the initial metasurface layer with 
transmission coefficients updated at every time slot, i.e., the transmission properties of its
meta-atoms are reconfigured at rate $f_{\text{s}} \eqdef 1/T_{\text{s}}$ (rapidly TV); 
(ii) a S-only block (in light-blue color in Fig.~\ref{fig:fig_1}), comprising the subsequent \(L-1\) 
metasurface layers with transmission coefficients fixed over each transmission interval of duration $T$, i.e., 
the response of such layers is reconfigured at rate $1/T$ (slowly TV). 
{\em It is noteworthy that the rate of reconfigurability of the first ST layer of the SIM is 
$M$ times larger than that of the other S-only layers}.
This two-timescale layered architecture enables joint spatial and temporal wavefront processing across the metasurface stack.
The transmitting meta-atoms of the TV layer are time-modulated under the adiabatic  condition 
$f_{\text{s}} \ll f_0$ \cite{Minkov_2017}.
In contrast to \cite{Li-2024,Li-2025,Li-2024-hw}, the present work focuses on ST-SIM architectures for inducing artificial channel fluctuations and enabling opportunistic multiuser scheduling under partial CSIT, while assuming ideal hardware in order to isolate the fundamental wave-domain effects.

Regarding the meta-atom type, the $L$ metasurface layers of the SIM can be further partitioned into two groups: the boundary layers ($1$ and $L$) and the $L-2$ intermediate layers ($2,\ldots,L-1$).
Each of the intermediate \( L-2 \) layers consists of \( Q \triangleq Q_x \times Q_y \) \emph{transmitting} meta-atoms (in green color in Fig.~\ref{fig:fig_1}) arranged in a rectangular grid, with \( Q_x \) and \( Q_y \) elements along the \( x \)- and \( y \)-axes, respectively, and inter-element spacing \( d_\text{meta} \). 
Hereinafter, the inter-element spacing of each metasurface layer $d_{\text{meta}}$ is assumed, for simplicity, to be equal to the inter-antenna spacing of the UPA $d_{\text{upa}}$, i.e., $d_{\text{upa}} = d_{\text{meta}}$.
By contrast, the boundary layers contain fewer than $Q$ transmitting meta-atoms. 
The first ST layer comprises $Z \triangleq Z_x \times Z_y \leq Q$ transmitting meta-atoms spatially aligned with the UPA 
grid ($Z \geq N$), with inter-element spacing $d_{\text{meta}}$; the remaining $Q - Z$ meta-atoms are \emph{perfectly absorbing} and surround the transmitting region (in red color in Fig.~\ref{fig:fig_1}). 
Likewise, the terminal layer $L$ comprises $V \triangleq V_x \times V_y \leq Q$ transmitting meta-atoms aligned with the same grid ($V \geq N$), with spacing $d_{\text{meta}}$, 
while the remaining $Q - V$ elements are perfectly absorbing. 
These layers are designed to decouple the dimensionality of the ST-SIM from 
the number $Q$ of transmitting meta-atoms. 
In order to highlight this property, we refer to the first and last metasurfaces as  \emph{dimensional adaptation layers} (DALs).

\begin{table*}[t]
\caption{Comparison between a conventional SIM and the proposed DAL-aided ST-SIM.}
\label{tab:sim_comparison}
\centering
\renewcommand{\arraystretch}{1.2}
\begin{tabular}{p{0.22\textwidth} p{0.35\textwidth} p{0.35\textwidth}}
\hline
\textbf{Characteristic} & \textbf{Conventional SIM} & \textbf{Proposed DAL-aided ST-SIM}\\
\hline
SIM modulation 
& Space-only (fixed over the coherence interval $T$) 
& Space-time (ST layer varies across time slots of duration $T_s$, with $T=M T_s$)\\[0.35em]

Boundary layers 
& All transmitting meta-atoms 
& DALs include transmitting and absorbing meta-atoms: the first ST-DAL has $Z$ transmitting and $Q-Z$ absorbing meta-atoms, 
while the last S-only DAL has $V$ transmitting and $Q-V$ absorbing meta-atoms\\[0.35em]

Intermediate layers 
& $Q$ transmitting meta-atoms 
& Same: $Q$ transmitting meta-atoms\\[0.35em]

Response dimension 
& Tied to $Q$: $\bm G\in\mathbb{C}^{Q\times N}$
& Decoupled from $Q$: $\bm G(t)\in\mathbb{C}^{V\times N}$\\[0.35em]

Degrees of control 
& Mainly via $L$ 
& Via both $L$ and $Q$\\
\hline
\end{tabular}
\end{table*}

In a conventional SIM architecture with dimensions $Q \times N$, the number of variables to be optimized is $Q N$, while 
the degrees of freedom at the designer's disposal are limited to the number of metasurface layers $L$. Consequently, the 
designer is constrained to increasing $L$ to enhance performance, which may adversely affect the convergence rate of the 
iterative algorithms employed to synthesize the transmission SIM coefficients, due to error propagation across layers.
In contrast, the proposed SIM architecture decouples the  number of transmitting antennas $N$ and the number of meta-atoms \( Q \) of the intermediate $L-2$ layers from the number of elements of the overall SIM response matrix, having dimensions $V \times Z$, thereby introducing \( Q \) as an independent design variable that can be optimized as the number of layers \( L \).
It is worth noting that the inclusion of the DAL is applicable to any SIM design, which is not limited to ST implementations only.

When an incident EM wave impinges on a meta-atom, the transmitted wave’s amplitude and phase are determined by the product of the incident field and the meta-atom's complex-valued transmission coefficient. The re-radiated wave then serves as a secondary source illuminating the subsequent layer, following the Huygens–Fresnel principle \cite{Goodman}. 
On the contrary, each wave striking one of the absorbing meta-atoms in the DALs is completely dissipated by integrated circuitry, preventing further propagation. 
From a hardware perspective, a strongly attenuating (almost absorbing) response can be realized using purely passive lossy meta-atoms. In our architecture, however, some layers already employ active meta-atoms with integrated amplifier chips to enable programmable amplitude control. 
Under 
field programmable gate array (FPGA) control, these meta-atoms operate as active artificial neurons, enabling a large dynamic modulation range 
(e.g., from $-22$ to $13$\,dB in \cite{Liu.2022}).
Therefore, these same active elements can also be driven into a highly lossy operating point,  where they effectively behave as near-perfect absorbers, strongly suppressing the propagation of the incident wave to subsequent layers.
While perfect absorption over a wide bandwidth is difficult to achieve \cite{Tsitsas.2017,Wang.2023}, moderate deviations from ideal absorption primarily result in small 
reflections that can be interpreted as perturbations of the effective transmission coefficients of the metasurface layers. 
Such effects do not alter the operation of the proposed synthesis algorithm, which optimizes the metasurface transmission 
coefficients to approximate the desired transfer matrix.
Hereinafter, for modeling convenience, the meta-atoms in the DALs will be modeled 
as perfect absorbers, which fully dissipate the incident energy.

Table~\ref{tab:sim_comparison} summarizes the key structural differences with respect to conventional SIM architectures.

\subsection{Wave propagation model through the ST-SIM}

In the following, we will denote with
$(n_x, n_y)$ the two-dimensional position of a generic UPA antenna element, where
$n_x \in \{0, 1, \ldots, N_x - 1\}$ and $n_y \in \{0, 1, \ldots, N_y - 1\}$, while
\be
n \eqdef n_x N_y + n_y \in \mathcal{N} \eqdef \{0,1,\ldots,N-1\}
\ee
represents its corresponding one-dimensional index.
Similarly, each transmitting meta-atom of the intermediate $L-2$ metasurface layers is located at $(q_x, q_y)$ with
$q_x \in \{0, 1, \ldots, Q_x - 1\}$ and $q_y \in \{0, 1, \ldots, Q_y - 1\}$, and is indexed as
\be
q \eqdef q_x Q_y + q_y \in \mathcal{Q} \eqdef \{0,1,\ldots,Q-1\}.
\ee
The same indexing applies to the $V$ transmitting meta-atoms of the terminal DAL, where each element
is located at $(v_x, v_y)$ with $v_x \in \{0, 1, \ldots, V_x - 1\}$ and
$v_y \in \{0, 1, \ldots, V_y - 1\}$, and is indexed as
\be
v \eqdef v_x V_y + v_y \in \mathcal{V} \eqdef \{0,1,\ldots,V-1\}.
\ee

We assume that the S-only block of the ST-SIM, inclu\-ding the layers $\ell \in \mathcal{L}^{\text{(s)}} \eqdef 
\{2,3,\ldots,L\},$ is implemented using
\emph{amplitude-controlled (AC)} and/or \emph{phase-controlled (PC)} metasurfaces~\cite{Dar.2025}.
AC layers allow programmable amplitude control via active circuitry, whereas PC layers,
which are nearly passive, permit programmable phase shifts. We denote the corresponding
index sets with $\mathcal{L}^{\text{(s)}}_{\text{ac}}$ and $\mathcal{L}^{\text{(s)}}_{\text{pc}}$ such that
\[
\mathcal{L}^{\text{(s)}}_{\text{ac}} \cap \mathcal{L}^{\text{(s)}}_{\text{pc}} = \emptyset
\quad \text{and} \quad
\mathcal{L}^{\text{(s)}} 
= \mathcal{L}^{\text{(s)}}_{\text{ac}} \cup \mathcal{L}^{\text{(s)}}_{\text{pc}}.
\] 

Let $\gamma_{\ell,q} = \alpha_{\ell,q} \, e^{j \phi_{\ell,q}}$, for $q \in \mathcal{Q}$ and
$\ell \in \mathcal{L}^{\text{(s)}}-\{L\}$, denote the transmission coefficient of the
$q$-th meta-atom in the $\ell$-th layer of the S-only block (excluding the terminal DAL), where
$\alpha_{\ell,q}$ and $\phi_{\ell,q}$ are its amplitude and phase, respectively.
We define 
$\pmb{\gamma}_\ell \triangleq [\gamma_{\ell,0}, \ldots, \gamma_{\ell,Q-1}]^\top \in \Cset^Q$, 
which collects the transmission coefficients of all $Q$ meta-atoms in the $\ell$-th layer, and the diagonal matrix
$\bm{\Gamma}_\ell \triangleq \mathrm{diag}(\pmb{\gamma}_\ell)$.
\begin{table*}[t]
\scriptsize
\centering
\caption{Main system parameters.}
\label{tab:example-1}
\begin{tabular}{cc|cc}
\hline
\textbf{Symbol} & \textbf{Meaning} & \textbf{Symbol} & \textbf{Meaning} \\
\hline
$f_0$ & Carrier frequency & $N$ & Number of UPA transmit antennas \\
$\lambda_0$ & Carrier wavelength & $L$ & Number of metasurface layers \\
$d_{\text{upa}}$ & Spacing between adjacent UPA antennas &  $U$ & Number of system users \\
$d_{\text{meta}}$ & Spacing between adjacent meta-atoms & $Z$ & Number of meta-atoms of the input ST DAL \\
$s_\text{lay}$ & Spacing between adjacent layers of the ST-SIM & $Q$ & Number of meta-atoms in the intermediate S-only layers \\
$T$ & Channel coherence time &  $M$ & Number of time-slots per channel coherence interval  \\
$T_s$ & Duration of each time-slot ($T/M$) &  $V$ & Number of meta-atoms in the terminal S-only DAL \\
\hline
\end{tabular}
\end{table*}

In AC layers (i.e., for $\ell \in \mathcal{L}^{\text{(s)}}_{\text{ac}}$), the amplitudes $\alpha_{\ell,q}$
are software-controlled with a wide dynamic range (e.g., $\sim$35\,dB using dual amplifier
chips per meta-atom~\cite{Liu.2022}), while the phases
$\phi_{\ell,q} \equiv  \phi_{\text{ac}}^{(\ell,q)}$ are assumed known but not controllable.
In PC layers (i.e., for $\ell \in \mathcal{L}^{\text{(s)}}_{\text{pc}}$), the phases are digitally tunable,
whereas the amplitudes are not controllable and satisfy $\alpha_{\ell,q} = \alpha_{\text{pc}} \leq 1$ due to passivity
constraints~\cite{Dar.2025}.
With reference to the terminal S-only DAL (layer $L$), instead, only the $V$ transmitting meta-atoms have transmission coefficients
$\gamma_{L,v} = \alpha_{L,v} \,  e^{j \phi_{L,v}}$ for $v \in \mathcal{V}$, whereas the remaining
$Q - V$ absorbing meta-atoms have (ideally) zero amplitude response. 
In this case, we define 
$\pmb{\gamma}_L \triangleq [\gamma_{L,0}, \ldots, \gamma_{L,V-1}]^\top \in \Cset^V$
which collects the transmission coefficients of all $V$ meta-atoms in the terminal DAL, and the corresponding diagonal matrix
$\bm{\Gamma}_{\mathrm{dal}} \triangleq \mathrm{diag}(\pmb{\gamma}_L)$.

Regarding the ST block (i.e., layer~$1$), all meta-atoms are phase-controlled. Let $\delta_{z}(t)$ denote the time-varying transmission coefficient 
of the $z$-th meta-atom in the initial ST DAL, for $t \in [0,T)$, where the one-dimensional index $z$ is obtained from the 2-D coordinates $(z_x,z_y)$ as
\be
z \eqdef z_x \, Z_y + z_y \in \mathcal{Z} \eqdef \{0,1,\ldots,Z-1\}
\ee
with $z_x \in \{0,1,\ldots,Z_x-1\}$ and $z_y \in \{0,1,\ldots,Z_y-1\}$.
The base waveform $\delta_z(t)$ is defined over $M>0$ consecutive time-slots of duration
$T_{\text{s}} = T/M$ and is given by
\begin{equation}
\delta_{z}(t) = \sum_{m=0}^{M-1} \delta^{(m)}_{z}\, p(t - m \,T_{\text{s}})
\label{eq:Delta}
\end{equation}
where $p(t) = \Pi\!\left(\frac{t - T_{\text{s}}/2}{T_{\text{s}}}\right)$ represents a rectangular
pulse of duration $T_{\text{s}}$, and
$\delta^{(m)}_{z} = \beta \, e^{j \psi^{(m)}_z}$ is the transmission
coefficient of the $z$-th PC meta-atom in the $m$-th time slot, for
$m \in \mathcal{M} \eqdef \{0,1,\ldots,M-1\}$, with digitally controllable phase $\psi^{(m)}_z$
and fixed amplitude $\beta \leq 1$.
Let 
$\bm{\Delta}(t) \triangleq \mathrm{diag} \, [\pmb{\delta}(t)]$
denote the diagonal matrix whose main diagonal collects the rapidly TV coefficients
$\pmb{\delta}(t) \triangleq [\delta_0(t), \ldots, \delta_{Z-1}(t)]^\top \in \Cset^Z$
of the ST initial layer.

Regarding the wave propagation between successive layers, all forward-propagation matrices share the same elementwise 
expression derived from the Rayleigh–Sommerfeld diffraction propagator \cite{DiRenzo-ICC,Hanzo}:
\begin{equation}
\mathsf{K}(d;A,s) \triangleq \frac{A\,s}{2\pi d^{3}}\big(1-j\,\kappa_{0} \, d\big) \, e^{j\,\kappa_{0} \, d}
\label{eq:K}
\end{equation}
where \(\kappa_{0}\triangleq 2\pi/\lambda_{0}\) is the free-space wave number, with corresponding wavelength \(\lambda_{0}=c/f_{0}\) and light speed 
\(c=3\cdot10^{8}\) m/s.
\begin{table*}[t]
\centering
\caption{Dimensions of the main matrices in the ST–SIM model.}
\label{tab:matrix-dims}
\begin{tabularx}{\textwidth}{l @{\hspace{10em}} X @{\hspace{-5em}} c}
\hline
\textbf{Symbol} & \textbf{Description} & \textbf{Dimensions} \\
\hline
$\bm W_{1}$      & UPA $\to$ ST-DAL (layer 1)                                 & $Z\times N$ \\
$\bm W_{2}$      & ST DAL (layer 1) $\to$ layer 2                              & $Q\times Z$ \\
$\bm W_{\ell}$   & layer $\ell\!-\!1 \to$ layer $\ell$ \, (\text{for}  $\ell \in \{3,\ldots,L\!-\!1\}$)              & $Q\times Q$ \\
$\bm W_{L}$      & layer $L\!-\!1 \to$ S-only DAL (layer $L$)                       & $V\times Q$ \\
$\bm \Gamma_{\ell}$ & $\ell$-th TI layer coefficients ($\ell \in \{2,\ldots,L\!-\!1\}$)                & $Q\times Q$ \\
$\bm \Gamma_{\text{dal}}$ & S-only DAL coefficients                             & $V\times V$ \\
$\bm \Delta(t)$  & ST  DAL response (layer 1)                                         & $Z\times Z$ \\
$\bm G_{0}$      & S-only block response (ST  DAL output $\to$ S-only DAL output)                    & $V\times Z$ \\
$\bm G(t)$       & ST-SIM response (ST-DAL input $\to$ S-only DAL output)                                     & $V\times Z$ \\
$\widetilde{\bm G}(t) \eqdef \bm G(t)\bm W_{1}$ & UPA output $\to$ S-only DAL output                  & $V\times N$ \\
\hline
\end{tabularx}
\vspace{2pt}
\end{table*}
It models free-space EM wave propagation between two radiating elements separated by distance \(d\), with effective aperture area \(A\),  and inter-layer separation \(s\).
For a given metasurface, the transmission coefficients are typically obtained from unit-cell full-wave EM simulations under the assumption of local periodicity. This approximation is commonly adopted in metasurface modeling and is valid when higher-order grating modes are evanescent and sufficiently attenuated before reaching adjacent metasurface layers. Under these conditions, the interaction between layers can be accurately described through effective transmission coefficients and diffraction-based propagation models. 
Although the detailed EM implementation of the unit cells is beyond the scope of this work, the inter-element spacing and inter-layer separations considered in our examples 
in Section~\ref{sec:simul} are consistent with regimes in which the local periodicity assumption provides a good approximation of the inter-layer propagation.
Specifically, propagation between the UPA and the ST DAL (layer $1$) is described by the following matrix 
\be
\bm W_{1}\in\Cset^{Z\times N},\quad \text{with
$[\bm W_{1}]_{z,n}=\mathsf{K}\big(d_{z,n};A_{\text{bs}},s_{\text{bs}}\big)$}
\ee
where $A_{\text{bs}}$ is the effective area of the UPA antennas (evaluated at $f_0$), $s_\text{bs}$ denotes the 
spacing between the UPA and the first layer of the ST-SIM, and $d_{z,n}$ represents the distance between the $n$-th antenna of the UPA 
and the $z$-th meta-atom of the first layer and it is given by 
\be
d_{z,n} =  \sqrt{[(n_x-{z}_x)^2 + (n_y-{z}_y)^2] \, d^2_{\text{meta}}+s_{\text{bs}}^2}
\label{eq:dqq-1} \:.
\ee
Similarly, propagation between the ST DAL (layer $1$) and layer $2$ is described by
\be
\bm W_{2}\in\Cset^{Q\times Z},\quad \text{with 
$[\bm W_{2}]_{q,z}=\mathsf{K}\big(d_{q,z};A_{\text{meta}},s_{\text{lay}}\big)$}
\ee
where
\be
d_{q,z}=\sqrt{\big[(q_{x}-z_{x})^{2}+(q_{y}-z_{y})^{2}\big]\,d_{\text{meta}}^{2}+s_{\text{lay}}^{2}}
\ee
is the distance between the $z$-th element of layer $1$ and the $q$-th meta-atom of layer $2$.
For the intermediate layers, propagation from layer $\ell-1$ to layer $\ell$, with $\ell \in \{3,\ldots,L-1\}$, reads as 
\be
\bm W_{\ell}\in\Cset^{Q\times Q},\quad \text{with 
$[\bm W_{\ell}]_{\tilde q,q}=\mathsf{K}\big(d_{\tilde q,q};A_{\text{meta}},s_{\text{lay}}\big)$}
\ee
where $A_{\text{meta}}$ is the physical area of each meta-atom, while $d_{\tilde{q},q}$ represents the propagation distance between 
the $q$-th meta-atom of the $(\ell-1)$-th layer
and the $\tilde{q}$-th meta-atom of the $\ell$-th layer, whose expression is
\be
d_{\tilde{q},q} =  \sqrt{[(q_x-\tilde{q}_x)^2 + (q_y-\tilde{q}_y)^2] \, d^2_{\text{meta}}+s_\text{lay}^2} \: .
\ee

Finally, propagation between layer $L-1$ and the terminal S-only DAL (layer $L$) is described by
\be
\bm W_{L}\in\Cset^{V\times Q},\quad \text{with
$[\bm W_{L}]_{v,q}=\mathsf{K}\big(d_{v,q};A_{\text{meta}},s_{\text{lay}}\big)$}
\label{eq:WL}
\ee
where
\be
d_{v,q}=\sqrt{\big[(v_{x}-q_{x})^{2}+(v_{y}-q_{y})^{2}\big]\,d_{\text{meta}}^{2}+s_{\text{lay}}^{2}}
\label{eq:dist}
\ee
represents the distance between the $q$-th meta-atom of layer $L-1$ and the $v$-th transmitting meta-atom of the terminal DAL.
The end-to-end forward propagation TV matrix across the ST-SIM, from the input of the ST DAL to the output of the S-only DAL, is given by 
\begin{equation}
\bm G(t) \triangleq \bm G_0 \, \bm \Delta(t) \in \Cset^{V \times Z}
\label{eq:forward}
\end{equation}
for $t \in [0,T)$, where
\be
\bm G_0 \triangleq \bm \Gamma_{\mathrm{dal}} \, \bm W_{L} \, \bm \Gamma_{L-1} \, \bm W_{L-1} \cdots \bm \Gamma_{2} \, \bm W_{2} \in \mathbb{C}^{V \times Z}
\label{eq:G0}
\ee
models the forward propagation through the S-only multi-layer block, and
$\bm \Delta(t)$ 
is the TV diagonal response of the ST initial DAL.
For clarity, the main system parameters are summarized in Table~\ref{tab:example-1}, whereas 
descriptions and dimensions of all matrices characterizing the ST-SIM 
are listed in Table~\ref{tab:matrix-dims}.

\section{Signal models}
\label{sec:rx-signal}

\begin{figure*}[t]
\centering
\includegraphics[width=\linewidth]{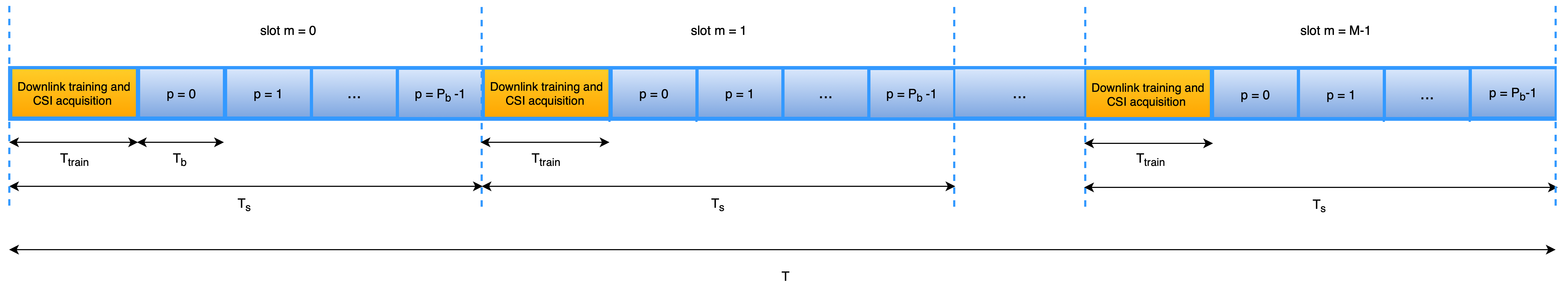}
\caption{Time-slot structure within a coherence block. Each slot of duration $T_{\text{s}}$ begins with a downlink training/CSI acquisition phase of length $T_{\text{train}}$, followed by $P$ data-symbol intervals of duration $T_{\text{b}}$ indexed by $p \in \{0,\ldots,P-1\}$. This pattern repeats $M$ times so that $M T_{\text{s}}$ spans the channel coherence time $T$.}
\label{fig:fig_2}
\end{figure*}

With reference to Fig.~\ref{fig:fig_2}, each one of the $M$ time slots comprises $P \triangleq P_{\text{t}} + P_{\text{b}}$ 
symbol intervals of duration $T_{\text{b}}$. The downlink training and CSI acquisition phase (described in the next section) occupies $P_{\text{t}}$ symbols, i.e., 
$T_{\text{train}} \eqdef P_{\text{t}} \, T_{\text{b}}$, followed by a payload of $P_{\text{b}}$ symbols, with $P_{\text{b}} > N$. Consequently,
\be
T_{\text{s}} = (P_{\text{t}} + P_{\text{b}}) \, T_{\text{b}} = T_{\text{train}} + P_{\text{b}} \, T_{\text{b}}.
\ee

The complex envelope of the narrowband continuous-time signal associated to the $n$-th data-stream ($n \in \mathcal{N}$) transmitted by the UPA is given by
\be
d_n(t) = \sqrt{\euscr{E}} \sum_{k=-\infty}^{+\infty} b_n(k) \, q(t-k \, T_\text b)
\label{eq:xn}
\ee
for $t \in [0,T)$, where the data-streams $\{b_n(k)\}_{n \in \mathcal{N}}$, for $k \in \mathbb{Z}$, are mo\-de\-led as mutually independent sequences of zero-mean
unit-variance independent and identically distributed (i.i.d.) complex random variables, emitted with rate $1/T_\text b$, 
$\euscr{E}$ represents the transmit energy uniformly associated to each of the $N$ data-streams, 
and $q(t)$ is the unit-energy square-root Nyquist pulse-shaping filter.

Let $\bm d(t) \triangleq [d_0(t), d_1(t), \ldots, d_{N-1}(t)]^\top \in \Cset^N$ denote the
complex baseband signal vector transmitted by the UPA. 
The signal impinging on the SIM is given by
\be
\bm x(t) \triangleq \bm W_1 \, \bm d(t)
= [\,x_0(t), x_1(t), \ldots, x_{Z-1}(t)\,]^\top \in \Cset^{Z}.
\label{eq:x}
\ee
After interacting with the ST layer and
passing through the multilayer structure, the SIM re-radiates the baseband vector
\be
\bm z(t) \triangleq \bm G(t)\,\bm x(t) = \sum_{z=0}^{Z-1} \bm g_z(t)\, x_z(t) \in \Cset^{V}
\label{eq:z}
\ee
where $\bm g_z(t) \in \Cset^{V}$ is the $z$-th column entry of the matrix $\bm G(t) =  [\bm g_0(t), \bm g_1(t), \ldots, \bm g_{Z-1}(t)]$.
Substituting \eqref{eq:x} into \eqref{eq:z} yields the input-output relationship 
\begin{equation}
\bm z(t)
= \widetilde{\bm G}(t)\,\bm d(t)
= \sum_{n=0}^{N-1} \widetilde{\bm g}_{n}(t)\, d_{n}(t)
\in \Cset^{V}
\label{eq:z-new}
\end{equation}
where 
\be
\widetilde{\bm G}(t) \triangleq \bm G(t) \, \bm W_{1}
= \big[\,\widetilde{\bm g}_{0}(t),\ldots,\widetilde{\bm g}_{N-1}(t)\,\big] \in \Cset^{V\times N}
\label{eq:overall}
\ee
denotes the {\em overall} ST-SIM response between the UPA output and the DAL output, and 
\be
\widetilde{\bm g}_{n}(t) \triangleq \bm G(t) \, \bm w_{1,n}
\ee
is the $n$th effective steering vector,
with $\bm w_{1,n}$ denoting the $n$th column of $\bm W_{1} = [\bm w_{1,0}, \bm w_{1,1}, \ldots, \bm w_{1,N-1}]$.
As explained in Section~\ref{sec:synthesis}, in each time slot, the ST-SIM independently generates a set
$\{\widetilde{\bm g}_n(t)\}_{n=0}^{N-1}$, uncorrelated with those used in previous slots.
Using \eqref{eq:xn}, \eqref{eq:x}, and \eqref{eq:z}, the {\em time–averaged} radiated power by the SIM is given by
\begin{align}
\euscr{P}_{\text{rad}}
&\triangleq \big\langle \Es\!\big[\|\bm z(t)\|^2\big] \big\rangle = \big\langle \Es\!\big[\|\widetilde{\bm G}(t)\,\bm d(t)\|^2\big] \big\rangle  \nonumber\\
&= \big\langle \Es\!\big[\|\bm G_{0}\,\bm\Delta(t)\,\bm W_{1}\,\bm d(t)\|^2\big] \big\rangle \nonumber\\
&= \euscr{E}\,\Big\langle
   \sum_{k\in\mathbb{Z}} q^{2}\!\big(t-kT_{\mathrm b}\big)
   \sum_{n=0}^{N-1} \Es\!\big[\|\bm G_{0}\,\operatorname{diag}(\bm w_{1,n})\,\pmb{\delta}(t)\|^2\big]
   \Big\rangle \nonumber\\
&= \euscr{P}_\text{sig}\, \frac{\beta^{2}}{N}\,
   \sum_{n=0}^{N-1}\sum_{z=0}^{Z-1} \big|[\bm W_{1}]_{z,n}\big|^{2}\,\|\bm g_{z}^{(0)}\|^{2}
\label{eq:Prad}
\end{align}
where $\langle \cdot \rangle$ denotes a time-average operator,  
$\|\cdot\|$ is the Frobenius matrix norm, 
$\bm G_0 = [\bm g_0^{(0)}, \ldots, \bm g_{Z-1}^{(0)}]$, and $\euscr{P}_\text{sig} \triangleq \tfrac{N\euscr{E}}{T_{\mathrm b}}$ 
denotes the average power of the input signal $\bm d(t)$. The last two equalities follow from the assumption that the data 
streams $\{d_n(t)\}_{n=0}^{N-1}$ are i.i.d. and the SIM 
transmission coefficients $\{\delta_n(t)\}_{n=0}^{N-1}$, together with the pairwise uncorrelated nature of their components.
Hereinafter, we will assume that no power amplification is realized by the SIM and, then, 
we will enforce the following constraint in its design:
\be
\|\bm g_z^{(0)} \|^2 = \frac{N}
{\beta^2 \, \|\bm W_1\|^2}  \:, \quad \text{for $z \in \mathcal{Z}$}
\label{eq:norm-constr}
\ee
which entails that the radiated power is equal to
$\euscr{P}_{\text{rad}} = \euscr{P}_\text{sig}$.
Strictly speaking, the AC layers in the S-only part of the SIM are used only to 
compensate for the propagation losses inside the SIM.
This choice differs from \cite{DiRenzo-ICC}, where the complex transmission coefficients of each layer exhibit unit amplitude implying, thus, that
each layer can be realized without utilizing resistive components (\emph{local design}).
We, instead, enlarge the set of feasible solutions of our design problem by allowing that the {\em globally} radiated power is preserved \cite{Dar.2025}.

At the receiver, which is assumed to lie in the SIM’s far field, the waveform is passed through a matched filter with impulse response $q(-t)$ and uniformly sampled at the symbol rate \(1/T_{\text{b}}\), under perfect timing synchronization.
Recalling that the ST-SIM operates in the adiabatic regime, one has
\be
q(-t) \star [\widetilde{\bm g}_n(t) \, d_n(t)] \approx \widetilde{\bm g}_n(t) \, [q(-t) \star d_n(t)]
\ee
where $\star$ stands for (linear) convolution,  which holds if $\widetilde{\bm G}(t)$ changes at a rate $f_\text s = 1/T_\text s$ much lower than the transmitted signal,
i.e., $f_\text s \ll 1/T_\text b$ (or, equivalently, $P \gg 1$).
Under this assumption, the resulting baseband signal $y_u(k) \triangleq y_u(k \, T_{\text{b}})$ received by the $u$-th user during the $k$-th symbol interval $[k \, T_{\text{b}},(k+1)\, T_{\text{b}})$, with $u \in \mathcal{U} \triangleq \{1,2,\ldots,U\}$ and $k \in \mathcal{K} \triangleq \bigcup_{m=0}^{M-1} \{\,mP+P_{\text{t}},\ldots,(m+1)P-1\,\}$, is given by
\begin{equation}
y_u(k)
= \sqrt{\euscr{E}}\,\sqrt{\varrho_u}\,
\sum_{n=0}^{N-1} \bm h_u^{H}\, \widetilde{\bm g}_n(k)\, b_n(k) + w_u(k) \:. 
\label{eq:yk}
\end{equation}
Here, the discrete-time low-pass equivalent response $\bm h_u \in \Cset^{V}$ models the frequency-flat block-fading channel from the ST-SIM to the $u$-th user, with $\mathbb{E}[\|\bm h_u\|^2]=1$ and coherence time $T$. The scalar factor
\begin{equation}
\varrho_u = \Big(\tfrac{\lambda_0}{4\pi d_0}\Big)^{\!2}\!\Big(\tfrac{d_0}{d_u}\Big)^{\!\eta}
\end{equation}
is the path loss of the $u$-th link, where $d_u$ denotes the SIM–to–user distance, $d_0$ is the far-field reference distance, and $\eta$ is the path-loss exponent. Moreover, $\bm g_n(k) \triangleq \bm g_n(kT_{\text{b}})$ is the sampled counterpart, at rate $1/T_{\text{b}}$, of the $n$-th PTV steering vector $\bm g_n(t)$, and $w_u(k) \triangleq w_u(kT_{\text{b}})$ is modeled as i.i.d. circularly symmetric complex Gaussian noise with zero mean and variance $\sigma_{w}^{2}$, statistically independent of $b_n(k)$ and $\bm h_u$ for any $n \in \mathcal{N}$, $k \in \mathbb{Z}$, and $u \in \mathcal{U}$.
It is worth noting that the steering vector $\bm g_n(k)$ varies across time-slots, while remaining constant within each slot, i.e., for  $k \in \mathcal{K}$. 

The adopted channel model assumes frequency-flat block fading with path-loss-dependent variance. 
This simplified representation is intentionally used to isolate the effect of the artificial channel fluctuations introduced by the proposed ST-SIM architecture. In particular, the objective of this work is to evaluate how ST-SIM randomization enables opportunistic multiuser scheduling under limited CSIT, independently of propagation-induced randomness.
More detailed propagation models including sparse multipath components, blockage effects, and spatial correlation across large apertures can be incorporated without modifying the proposed beamforming and scheduling framework. Likewise, recent studies have investigated SIMs in wideband communication systems, particularly 
MIMO systems based on orthogonal frequency-division multiplexing architectures where metasurface layers are optimized to jointly fit the frequency-selective 
channel response across subcarriers \cite{Li.2026,Li.2026-IM}. 
The analysis presented here therefore focuses on the fundamental behavior of the ST-SIM mechanism rather than on environment-specific propagation characteristics.

\section{User scheduling with partial CSIT}
\label{sec:random-ST-SIM}

We present the user scheduling strategy based on partial CSIT, which plays a central role in reducing feedback overhead.
Many recent beamforming techniques 
proposed for SIM-assisted multiuser beamforming (see, e.g., \cite{DiRenzo-ICC,Liu.2024,An_ArXiv_2025,Li.2025,Dar.2025}) 
rely on accurate estimation of the full CSIT for all users. 
When \emph{user selection} is performed at the transmitter, the BS selects the $N \leq U$ users exhibiting the highest signal-to-interference-plus-noise ratio (SINR), based on the acquired CSIT, and constructs the beamforming vectors by treating the estimated CSIT as perfect.
If channel reciprocity holds, BS can acquire the downlink CSI using pilot symbols sent by the users. 
Since the number of RF chains $N$ is smaller than the number of meta-atoms $V$ of the final metasurface layer, all the channels associated with $U$ users can be estimated by using at least $\lceil \tfrac{U \, V}{N} \rceil$ symbols~\cite{Yao.2024}, but this task
becomes increasingly cumbersome as the number of meta-atoms $V$ grows.
By contrast, when channel reciprocity does not hold, a closed-loop CSIT estimation approach must be used~\cite{Caire}. In this case, the $u$-th user (with $u \in \mathcal{U}$) estimates its channel vector $\bm{h}_u \in \mathbb{C}^V$ based on downlink training symbols, and then feeds back the estimate $\bm{h}_u^{\text{(est)}}$ to the BS.

CSIT acquisition and feedback pose a significant challenge in SIM-based beamforming architectures.
One key difficulty in CSIT acquisition stems from the fact that the BS typically has fewer RF chains 
than the number of meta-atoms in the final metasurface layer, which dictates the effective dimensionality of the channel to be estimated. 
Another challenge concerns CSIT feedback. Whether analog feedback is used, where each user transmits unquantized channel coefficients via the real and imaginary parts of a complex modulation symbol~\cite{Marzetta}, or digital feedback is employed, where each user's estimated channel vector \( \bm{h}_u^{\text{(est)}} \) is quantized and fed back to the BS, the resulting feedback overhead scales with \( U \, V \) within each channel coherence interval. 
In both cases, this overhead may become costly in systems with rate-limited feedback channels, particularly in scenarios with a large number of users and/or densely deployed SIM elements.
To address these limitations, the proposed ST-SIM architecture implements a 
beamforming strategy in the wave domain that relies on {\em partial} CSIT and opportunistically serves the users. 

For $u \in \mathcal{U}$ and $k \in \mathcal{K}$, the complex baseband signal in \eqref{eq:yk} received during the $k$-th symbol interval $[kT_{\text b},(k+1)T_{\text b})$ by the $u$-th user can be rewritten as
\begin{equation}
y_u(k) = \sqrt{\euscr{E}} \sum_{n=0}^{N-1} c_{u,n}(k)\, b_n(k) + w_u(k)
\label{eq:sig-rx}
\end{equation}
where the equivalent time-varying scalar channel
\be
c_{u,n}(k) \triangleq \sqrt{\varrho_u}\,\bm h_u^{H}\, 
\widetilde{\bm g}_n(k)
\ee
is equal to the projection of the $u$-th channel vector $\sqrt{\varrho_u}\,\bm h_u$ onto the $n$-th steering vector $\widetilde{\bm g}_n(k)$.
We recall that each time-slot begins with a downlink training phase of duration $T_\text{train} = P_\text t \, T_\text b$ (see Fig.~\ref{fig:fig_2}), where 
$P_\text t \ge N$. 
During the channel estimation phase corresponding to the $m$-th time-slot, the $u$-th user can acquire  
the channel vector 
\be
\widetilde{\bm c}_u(m) \triangleq [\widetilde{c}_{u,0}(m),\, \widetilde{c}_{u,1}(m),\, \dots,\, \widetilde{c}_{u,N-1}(m)]^\mathsf{T} \in \Cset^{N}
\ee
with $\widetilde{c}_{u,n}(m) \eqdef c_{u,n}(m P)$, by using standard training-based estimation techniques \cite{Kay-book}.

After acquiring CSI, each user {\em locally} computes the index $n^\star_{u,m}$ maximizing its SINR as follows
\be
n^\star_{u,m} \eqdef \arg \max_{\substack{n \in \mathcal{N}}} \, \, \text{SINR}_{u,n}(m)
\label{eq:sinr}
\ee
where
\be
\text{SINR}_{u,n}(m) \eqdef \frac{|\widetilde{c}_{u,n}(m)|^2}{\displaystyle\sum_{\substack{j \in \mathcal{N} \\ j \ne n}} |\widetilde{c}_{u,j}(m)|^2 + \frac{\sigma_w^2}{\euscr{E}}}
\ee
and feeds the information $\text{SINR}_{u,n^\star_{u,m}}(m)$ back to the BS, for each time-slot.
Based on the partial information provided by the $U$ users in the network, the base station opportunistically schedules transmissions in every time-slot to the 
$N$ users who exhibit the highest
$\text{SINR}_{u,n^\star_{u,m}}(m)$ values, where the maximization is performed over $u$ for each steering vector index $n \in \mathcal{N}$.
Since each user feeds back only its preferred beam index \(n_{u,m}^{\star}\) and the corresponding SINR value, each user can be associated with at most one beam in a given time-slot. Hence, distinct user assignment across beam indices is enforced by construction. The BS then groups the received feedback according to the reported beam index and, for each \(n\in\mathcal{N}\), selects the user with the largest reported SINR among those declaring \(n=n_{u,m}^{\star}\). If no user reports beam index \(n\), that beam is left inactive during the corresponding slot. Therefore, fewer than \(N\) users may be served in a slot, but no user is assigned to multiple beams simultaneously.
In this manner, the system can opportunistically serve up to $N \, M$ different users in each time interval $[0,T)$.
The considered scheduling protocol assumes that, in each time slot, every user feeds back the index of its 
preferred steering vector together with a scalar quality metric (e.g., the corresponding SINR). 
Thus, the aggregate feedback overhead scales linearly with both the number of users and the number of time slots per coherence interval. 
Several lightweight reporting mechanisms can be adopted to further reduce the feedback load without modifying the proposed scheduling principle
\cite{Vis-2002,Love2008}. 
For instance, users may report quantized versions of the SINR metric, employ threshold-based reporting (where feedback is transmitted only if 
the best-beam SINR exceeds a predefined threshold), or adopt contention-based or group-based feedback protocols suitable for dense-user scenarios. 

We are now in the position to define the user selected for each steering vector index. 
For a given time-slot $m\in\mathcal{M}$ and beam index $n\in\mathcal{N}$, 
let $\mathcal{U}_n(m)$ denote the set of users that reported $n=n_{u,m}^\star$. 
If $\mathcal{U}_n(m)\neq\emptyset$, the BS selects the user
\begin{equation}
u_n^\star(m) \triangleq \arg\max_{u\in\mathcal{U}_n(m)} \text{SINR}_{u,n}(m).
\end{equation}
We denote by $\mathcal{U}^\star(m)$ the set of scheduled users in the $m$-th time-slot, i.e.,
$\mathcal{U}^\star(m)\subseteq\mathcal{U}$, with $|\mathcal{U}^\star(m)|\le N$.
If $\mathcal{U}_n(m)=\emptyset$, the corresponding beam $n$ remains inactive during that slot. 
For notational convenience, when a user is selected for beam $n$ within time-slot $m$, 
we denote the corresponding SINR as
$\text{SINR}^\star_{u_n^\star,n}(m)$.
Given the received signal model \eqref{eq:sig-rx}, the achievable rate for user ${u}^\star_n$ in the $m$-th time-slot is 
\be
\rate_{{u}^\star_n}(m) = \log_2 \left( 1 + {\text{SINR}}^\star_{{u}^\star_{{n}}}(m) \right)
\label{eq:R}
\ee
where
\be
{\text{SINR}}^\star_{{u}^\star_{{n}}}(m) \eqdef \frac{|\widetilde{c}_{{u}^\star_{{n}},n}(m)|^2}{\displaystyle\sum_{\substack{j \in {\mathcal{N}} \\ j \ne n}} |\widetilde{c}_{{u}^\star_{{n}},j}(m)|^2 + \frac{\sigma_w^2}{\euscr{E}}} 
\ee
with $\widetilde{c}_{{u}^\star_{{n}},j}(m) = \sqrt{\varrho_{{u}^\star_{{n}}}} \, \bm h^H_{{u}^\star_{{n}}}  \, \widetilde{\bm g}_j(m)$ denoting the effective channel coefficient between the ST-SIM and the selected user ${u}^\star_{{n}}$ associated with steering vector $\widetilde{\bm g}_j(m)$.
The overall performance of the multiuser communication system can be evaluated in terms of the time-averaged (TA) sum rate, computed over the $M$ time-slots, as
\be
\rate = \frac{1}{M} \, \sum_{m=0}^{M-1} \, \sum_{{u^\star_n \in \mathcal{U}^\star}} \rate_{{u}^\star_n}(m) 
\label{eq:rate}
\ee
where $\rate_{{u}^\star_n}(m)$ denotes the achievable rate for user ${u}^\star_n$ in the $m$-th time-slot and is given by \eqref{eq:R}.
The {\em effective} sum-rate accounts for the training and CSI-feedback overhead within each channel coherence interval. 
Specifically, we define
\begin{equation}
\rate_{\mathrm{eff}} \triangleq \xi \, \rate 
\label{eq:Reff}
\end{equation}
where $0< \xi \le 1$ denotes the fraction of symbols available for data transmission. 
The value of $\xi$ depends on the adopted transmission scheme, as well as 
on the associated training and CSI-feedback procedures.

The following three remarks highlight that the effective number of time-slots \( M \) must be carefully balanced 
to trade off between maximizing multiuser diversity and minimizing both training and CSI feedback overhead. 

\vspace{1mm}
{\em Remark~1:}
Unlike conventional 
S-only SIM schemes,
the time modulation employed in the proposed ST-SIM architecture enables the exploitation of \emph{multiuser diversity}. When the channel between the SIM and the users is \emph{slowly time-varying}, with a coherence time \( T \) spanning 
multiple symbol periods (i.e., \( T = M \, T_s = M \, P  \, T_\text b \), with \( M \, P \gg 1 \)), the channel remains nearly constant 
over several transmission intervals. In such a scenario, only \( N \) out of the $U$ users
can be scheduled and effectively served within each coherence interval. As the total number of users \( U \) increases, 
this  leads to significant unfairness across the network~\cite{Tse-book}.
In contrast, the proposed ST-SIM-based beamforming approach introduces \emph{artificial time variations} into the wireless channel via time modulation. This temporal diversity allows the system to opportunistically schedule up to \( M N \) users for transmission. Consequently, the ST-SIM architecture improves fairness and enables the exploitation of multiuser diversity, even under slow channel dynamics.

\vspace{1mm}
{\em Remark~2:}
The proposed partial-CSI-based scheme requires the transmission of at least $N$ pilot symbols per time-slot. As a result, the total number of training symbols transmitted within each channel coherence interval is given by
\be
O_{\text{train}}^{\text{(part)}} = NM \: .
\ee
By contrast, full-CSI-based schemes require the transmission of 
\(
O_{\text{train}}^{\text{(full)}} = V
\)
training symbols per coherence interval under a time division duplex (TDD) protocol.
To ensure that the training overhead of the proposed scheme does not exceed that of full-CSI-based strategies, i.e., \( O_{\text{train}}^{\text{(part)}} \leq O_{\text{train}}^{\text{(full)}} \), the number of time-slots \( M \) should satisfy the following condition:
\be
0 < M \leq \frac{V}{N} 
\label{eq:rel}
\ee
which is typically easy to satisfy in practice, given the large number of meta-atoms in the SIM architecture.

\vspace{1mm}
{\em Remark~3:}
In the proposed partial-CSI-based beamforming scheme, each user feeds back a scalar value to the BS in every time-slot. Consequently, the associated CSI feedback overhead is given by
\be
O_{\text{feed}}^{\text{(part)}} = \eta \, U M
\ee
where the proportionality factor \( \eta \) depends on whether analog or digital feedback is employed. This overhead scales with both the number of users \( U \) and the number of time-slots \( M \), differently from full-CSI-based schemes, where the feedback overhead scales as \( U \, V \), and 
may become non-negligible even for moderate values of $V$.

\section{Transmission Coefficient Synthesis for randomized ST-SIM}
\label{sec:synthesis}

At the beginning of each time slot, the ST-SIM has to generate a set of \( N \)  
beamforming vectors \( \{\widetilde{\pmb{g}}_n(t)\}_{n=0}^{N-1} \), each used to modulate the corresponding information signal \( d_n(t) \).
Recalling the structure of the overall forward propagation matrix in~\eqref{eq:forward}, the generation of these steering vectors involves two distinct steps. 
The first step consists of generating a {\em fixed} matrix $\bm G_0$, 
which models the forward propagation through the S-only multilayer block of the SIM.
This step is performed once, at the beginning of each channel coherence interval of duration $T$. 
The second step, in contrast, is repeated at the beginning of each time-slot of duration $T_\text{s}$ and consists of 
{\em randomly} 
generating the rapidly TV transmission coefficients $\pmb{\delta}(t)$ of the initial ST DAL.

{\em Step 1 - Generation of the matrix $\bm G_0$: } Let $\bm{G}_{\text{targ}} \in \mathbb{C}^{V \times Z}$ denote the 
target matrix that models the forward propagation through the S-only multilayer block of the SIM, namely, 
from the output of the ST DAL to the output of the S-only DAL. It can be expressed as
\begin{equation}
\bm{G}_{\text{targ}} \triangleq \left[\, \bm{g}_{\text{targ},0},\, \bm{g}_{\text{targ},1},\, \ldots,\, \bm{g}_{\text{targ},Z-1} \,\right]
\end{equation}
where $\bm{g}_{\text{targ},n} \in \mathbb{C}^V$ denotes the desired steering vector associated with the $n$-th transmitted stream.
The optimization framework does not require any particular structure of $\bm{G}_{\text{targ}}$, whose columns only need 
to satisfy the norm constraint \eqref{eq:norm-constr}.

Our objective is to synthesize the transmission coefficients 
$\{\pmb{\gamma}_\ell\}_{\ell \in \mathcal{L}^{(\text{s})}}$ 
of the S-only metasurface layers such that the matrix $\bm{G}_0$ in \eqref{eq:forward} closely approximates $\bm{G}_{\text{targ}}$.
This leads to the constrained least-squares (LS) optimization problem
\begin{equation}
\min_{\{\pmb{\gamma}_\ell\}_{\ell \in \mathcal{L}^{(\text{s})}}} \quad f\left(\{ \pmb{\gamma}_\ell \}_{\ell \in \mathcal{L}^{\text{(s)}}}\right)
\end{equation}
where the objective function is defined as
\begin{equation}
f\left(\{ \pmb{\gamma}_{\ell \in \mathcal{L}^{\text{(s)}}} \}\right) \triangleq \|\bm{G}_0 - \bm{G}_{\text{targ}}\|^2
\label{eq:obj}
\end{equation}
under the following constraints:

\begin{align}
& \bm G_0 \triangleq 
  \bm \Gamma_{\text{dal}} \, 
  \bm W_{L} \, 
  \bm \Gamma_{L-1} \, 
  \bm W_{L-1} \cdots 
  \bm \Gamma_{2} \, 
  \bm W_{2} 
\label{eq:c1} \\
&   \|\bm g_z^{(0)} \|^2 = \frac{N}
{\beta^2 \, \|\bm W_1\|^2}, \quad z \in \mathcal{Z}
\label{eq:c2-bis} \\
& \bm{\Gamma}_\ell = \mathrm{diag}(\pmb{\gamma}_\ell), 
  \quad \ell \in \mathcal{L}^{(\text{s})}-\{L\}
\label{eq:c3} \\
& \bm{\Gamma}_\text{dal} = \mathrm{diag}(\pmb{\gamma}_\text{dal}) \\
& \gamma_{\ell,s} = 
\begin{cases}
\alpha_{\text{pc}} \, e^{j \phi_{\ell, s}}, 
  & \ell \in \mathcal{L}^{(\text{s})}_{\text{pc}} \\[4pt]
\alpha_{\ell, s} \, e^{j \phi_{\text{ac}}^{(\ell, s)}}, 
  & \ell \in \mathcal{L}^{(\text{s})}_{\text{ac}}
\end{cases}
\label{eq:c4} \\
& \alpha_{\text{min}} \leq \alpha_{\ell, s} \leq \alpha_{\text{max}}, 
  \quad \ell \in \mathcal{L}^{(\text{s})}_{\text{ac}} 
\label{eq:c6} 
\end{align}
with $s \in \mathcal{S} \eqdef \{0,1,\ldots,S-1\}$, where 
$\mathcal{S}=\mathcal{Q}$ (thus, $S=Q$) when $\ell \neq L$, 
and $\mathcal{S}=\mathcal{V}$ (thus, $S=V$) when $\ell = L$.
Inequality~\eqref{eq:c6} enforces the so-called {\em amplitude constraint}~\cite{Dar.2025}, which applies specifically 
to amplitude-controlled layers. This constraint reflects the practical limitation that meta-atoms in AC layers modulate 
the amplitude of the incident wave by adjusting the voltage supplied by the embedded amplifier circuits. Due to the physical 
characteristics of these chips, the achievable amplitudes are bounded within a finite range, determined by the allowable 
supply voltage. As a result, the corresponding transmission coefficients must lie within a prescribed interval \([ \alpha_{\text{min}}, \alpha_{\text{max}} ]\).

The above problem is solved using a projected gradient descent (PGD) algorithm~\cite{Beck}, which iteratively updates the transmission coefficients layer by layer while enforcing the corresponding amplitude and phase constraints.
At each iteration $\kappa$, the PGD algorithm performs the updates
\begin{align}
\pmb{\phi}^{(\kappa+1)}_\ell &= \pmb{\phi}^{(\kappa)}_\ell - \lambda^{(\kappa)}_{\pmb{\phi}_\ell} \nabla_{\pmb{\phi}^{(\kappa)}_\ell} f\left(\{ \pmb{\gamma}_{\ell \in \mathcal{L}^{\text{(s)}}} \}\right), \hspace{1.5mm} \ell \in \mathcal{L}^{\text{(s)}}_{\text{pc}} \label{eq:grad1} \\
\pmb{\alpha}^{(\kappa+1)}_\ell &= \mathcal{P}_A \left[ \pmb{\alpha}^{(\kappa)}_\ell - \lambda^{(\kappa)}_{\pmb{\alpha}_\ell} \nabla_{\pmb{\alpha}^{(\kappa)}_\ell} f\left(\{ \pmb{\gamma}_{\ell \in \mathcal{L}^{\text{(s)}}} \}\right) \right], \hspace{1.5mm} \ell \in \mathcal{L}^{\text{(s)}}_{\text{ac}}
\label{eq:grad2}
\end{align}
where $\pmb{\phi}_\ell \eqdef [\phi_{\ell,0}, \ldots, \phi_{\ell,S-1}]^\top$ and 
$\pmb{\alpha}_\ell \eqdef [\alpha_{\ell,0}, \ldots, \alpha_{\ell,S-1}]^\top$ denote the phase and 
amplitude vectors for layer $\ell$, respectively. The step sizes $\lambda^{(\kappa)}_{\pmb{\phi}_\ell}$ 
and $\lambda^{(\kappa)}_{\pmb{\alpha}_\ell}$ are determined using a backtracking line search \cite{Beck}.
The amplitude projection operator $\mathcal{P}_A[\alpha]$ is defined by the relations
\begin{align}
A &\triangleq \left\{\alpha \in \mathbb{R} \,:\, \alpha_{\text{min}} \leq \alpha \leq \alpha_{\text{max}} \right\} \\
\mathcal{P}_A[\alpha] &= \arg\min_{\widetilde{\alpha} \in A} |\widetilde{\alpha} - \alpha|.
\end{align}
The PGD iterations \eqref{eq:grad1}-\eqref{eq:grad2}
continue until a convergence criterion is met, either a sufficiently small variation in the cost function or a maximum number of iterations $\kappa_{\max}$ is reached. Upon convergence, the set of optimized transmission coefficients $\{\pmb{\gamma}^\star_\ell\}_{\ell \in \mathcal{L}^{(\text{s})}}$ configures the S-only part of the SIM.

To compute the gradients in \eqref{eq:grad1} and \eqref{eq:grad2}, we preliminary observe that \eqref{eq:obj} can be decomposed as
\begin{equation}
f\left(\{ \pmb{\gamma}_\ell \}\right) \eqdef \|\bm{G}_0 - \bm{G}_{\text{targ}}\|^2_F = \sum_{z=0}^{Z-1} \| \bm{g}_z^{(0)} - \bm{g}_{\text{targ},z} \|^2
\end{equation}
where we recall that 
$\bm{g}_z^{(0)}$ is the $z$-th column of $\bm{G}_0$, and $\bm{g}_{\text{targ},z}$ is the corresponding target steering vector.
According to \eqref{eq:c1}, each column $\bm{g}_z{(0)}$, for $z \in \mathcal{Z}$, admits the factorization
\begin{equation}
\bm{g}_z = \bm{E}_\ell \, \mathrm{diag}(\bm{b}_{\ell,z}) \, \pmb{\gamma}_\ell
\end{equation}
where, for $\ell \in \mathcal{L}^{\text{(s)}}$, $\bm{E}_\ell \in \mathbb{C}^{V \times Q}$ is extracted from $\bm G_0$ as $\bm{E}_\ell \triangleq \bm{\Gamma}_{\text{dal}} \, 
\bm{W}_L \, \bm{\Gamma}_{L-1} \cdots \bm{\Gamma}_{\ell+1} \, \bm{W}_{\ell+1} \in \mathbb{C}^{V\times Q}$, and $\bm{b}_{\ell,z}$ is the $z$-th column of  
$\bm{B}_\ell \triangleq \bm{W}_\ell \, \bm{\Gamma}_{\ell-1} \cdots \bm{\Gamma}_2 \, \bm{W}_2 \in \mathbb{C}^{Q\times Z}$.
Using this layered structure, the gradients of the objective function can be expressed as
\begin{align*}
\nabla_{\pmb{\phi}_\ell} f\left(\{ \pmb{\gamma}_\ell \}\right) 
&= 2\, \Im\left\{ \mathrm{diag}(\pmb{\gamma}^*_\ell) \left( \bm{A}_\ell \, \pmb{\gamma}_\ell - \bm{v}_\ell \right) \right\} \: , \quad \ell \in \mathcal{L}^{\text{(s)}}_{\text{pc}} \\
\nabla_{\pmb{\alpha}_\ell} f\left(\{ \pmb{\gamma}_\ell \}\right)
&= 2\, \Re\left\{ \mathrm{diag}(\pmb{\gamma}^*_\ell) \left( \bm{A}_\ell \, \pmb{\gamma}_\ell - \bm{v}_\ell \right) \right\} \: , \quad \ell \in \mathcal{L}^{\text{(s)}}_{\text{ac}}
\end{align*}
where the auxiliary matrices and vectors are defined as
\begin{align}
\bm{A}_\ell & \eqdef  \sum_{z=0}^{Z-1} \mathrm{diag}(\bm{b}^*_{\ell,z}) \, \bm{E}_\ell^\mathsf{H} \, \bm{E}_\ell \, \mathrm{diag}(\bm{b}_{\ell,z}) \nonumber \\
&= (\bm{B}_\ell^* \, \bm{B}_\ell^\top) \circ (\bm{E}_\ell^\mathsf{H} \, \bm{E}_\ell) \in \mathbb{C}^{Q \times Q}\\
\bm{v}_\ell & \eqdef  \sum_{z=0}^{Z-1} \mathrm{diag}(\bm{b}^*_{\ell,z})\,  \bm{E}_\ell^\mathsf{H} \, \bm{g}_{\text{targ},z}
= \left[ \bm{E}_\ell^\mathsf{H} \, \circ (\bm{B}_\ell^* \, \bm{G}_{\text{targ}}^\top) \right] \bm{1}_Q
\end{align}
with $\bm{1}_Q$ denoting the $Q \times 1$ all-ones vector, and $\circ$ being the Hadamard (element-wise) product.
The per-iteration complexi\-ty of the PGD synthesis is dominated by the computation of the matrices 
$\mathbf{E}_\ell^H\mathbf{E}_\ell$ and $\mathbf{B}_\ell^\ast\mathbf{B}_\ell^T$, yielding a complexity scaling of 
$\mathcal{O}\left((L-1) \, (V+Z) \, Q^2\right)$ per iteration.

{\em Step 2 - Generation of the matrix \( \bm{\Delta}(t) \):}  
The transmission coefficient \( \delta_z(t) \) of the \( z \)-th meta-atom in the initial 
DAL, with \( z \in \mathcal{Z} \), is generated according to~\eqref{eq:Delta}.
Specifically, at the beginning of each time-slot, 
we choose the digitally controllable phases $\psi^{(m)}_z$ of the transmission coefficients \(\delta^{(m)}_{z} = \alpha_{\text{dal}} \, e^{j \psi^{(m)}_z}\) in \eqref{eq:Delta}, for $m \in \mathcal{M}$ and $z \in \mathcal{Z}$, as a sequence of i.i.d. random variables with respect to both $m$ and $z$, where each random variable 
$\psi^{(m)}_z$ is uniformly distributed in the interval $[0,2 \pi)$.
By varying the phases of the meta-atoms in the first layer across both space and time, the SIM implements a ST beamforming $\widetilde{\bm G}(t)$ that randomly generates \( N \) steering vectors. At this point, relying only on partial CSIT, the transmitter schedules the transmission towards the \( N \) users that are closest to the resulting beams, which are more likely to ensure signal power maximization as the total number of users \( U \) increases. This fact is corroborated by the numerical results shown in the next section.

Algorithm~\ref{alg:sim_synthesis_randomization} summarizes the pseudo-code of the DAL-aided ST-SIM synthesis procedure.
It is important to distinguish between the offline block-level synthesis cost and the online slot-level update cost. The former is dominated by the PGD 
design of the S-only block and is incurred once per coherence interval, whereas the latter only involves random phase refresh of the ST layer 
and simple user scheduling operations.

{\em Remark~4:}
The proposed ST-SIM-based randomization should not be confused with classical random beamforming techniques, such as those introduced in opportunistic beamforming 
frameworks \cite{Sharif-2005,Vis-2002}. In these schemes, randomness is introduced at the digital baseband level by generating random precoding or beamforming vectors, while the physical wireless channel remains invariant over each coherence interval. In contrast, the proposed ST-SIM framework performs randomization directly in the wave domain through a ST metasurface layer that physically modulates the EM wavefront. This mechanism induces artificial time variations in the effective channel itself, even under slowly varying propagation conditions, thereby enabling multiuser diversity through physically induced channel fluctuations rather than digital precoder randomness. As a result, opportunistic scheduling can be realized with partial CSIT and reduced reliance on 
high-dimensional digital beamforming architectures.

{\em Remark~5:}
The proposed ST-SIM-based randomization is fundamentally different from conventional multi-beam training or beam-sweeping techniques used for beam alignment and channel acquisition
\cite{Sharif-2005,Hur-2013}. Beam-sweeping methods sequentially transmit a predefined set of deterministic beams from a finite codebook with the primary aim of identifying favorable propagation directions and estimating channel parameters. In contrast, the proposed ST-SIM framework is not intended for beam discovery or alignment. Instead, it introduces randomized ST wave-domain transformations via a time-modulated metasurface layer, which physically induce artificial time variations in the effective channel. This mechanism enables opportunistic scheduling and multiuser diversity under partial CSIT, without relying on explicit beam indexing, deterministic beam codebooks, or beam selection procedures.

{\em Remark~6:}
Besides the adiabatic condition $f_s \ll f_0$, a practically relevant requirement is that the settling time of the ST layer be sufficiently shorter than the slot duration $T_s$, so that the metasurface reaches its new configuration before payload transmission in the corresponding slot. For instance, with $T=5$ ms and $M=4$, one has $T_s=1.25$ ms and hence $f_s=800$ Hz, i.e., a reconfiguration speed on the order of kHz. Therefore, the proposed scheme does not require symbol-rate switching, since the ST coefficients are updated only at slot boundaries. This requirement is compatible with fast electronically tunable implementations, such as PIN-diode- or varactor-based metasurfaces \cite{Zhang.2018}, whereas slower technologies such as liquid-crystal-based metasurfaces \cite{Schwarzbeck.2026} may require smaller values of $M$ or longer channel coherence times.

\begin{algorithm}[t]
\caption{Synthesis of the DAL-aided ST-SIM}
\label{alg:sim_synthesis_randomization}
\begin{algorithmic}[1]
\Require Target matrix $\bm G_{\mathrm{targ}}\in\mathbb{C}^{V\times Z}$; propagation matrices $\{\bm W_\ell\}$;
index sets $\mathcal{L}^{(s)}=\mathcal{L}^{(s)}_{\mathrm{pc}}\cup\mathcal{L}^{(s)}_{\mathrm{ac}}$;
amplitude bounds $[\alpha_{\min},\alpha_{\max}]$;
maximum number of iterations $\kappa_{\max}$; number of time-slots $M$.
\Ensure Time-invariant response $\bm G_0$ and time-varying randomization matrices $\{\bm\Delta[m]\}_{m=0}^{M-1}$.
\Statex \textit{\textbf{Step 1: Synthesis of $\bm G_0$ (performed once at the beginning of the algorithm)}}
\State Initialize transmission coefficients $\pmb{\gamma}_\ell^{(0)}$ for $\ell\in\mathcal{L}^{(s)}$
\For{$\kappa = 0,1,\ldots,\kappa_{\max}-1$}
  \For{$\ell\in\mathcal{L}^{(s)}$}
    \State $\bm E_\ell \triangleq \bm\Gamma_{\mathrm{dal}}\bm W_L \bm\Gamma_{L-1}\bm W_{L-1}\cdots
    \bm\Gamma_{\ell+1}\bm W_{\ell+1}$
    \State $\bm B_\ell \triangleq \bm W_\ell \bm\Gamma_{\ell-1}\bm W_{\ell-1}\cdots \bm\Gamma_2\bm W_2$
    \State $\bm A_\ell \triangleq (\bm B_\ell^\ast \bm B_\ell^\top)\circ(\bm E_\ell^{\mathrm H}\bm E_\ell)$
    \State $\bm v_\ell \triangleq \Big(\bm E_\ell^{\mathrm H}\circ(\bm B_\ell^\ast \bm G_{\mathrm{targ}}^\top)\Big)\bm 1_Q$
    \State Compute the difference $\bm r_\ell \triangleq \bm A_\ell \, \pmb{\gamma}_\ell-\bm v_\ell$        \If{$\ell\in\mathcal{L}^{(s)}_{\mathrm{pc}}$}
    \State Compute
    $\nabla_{\pmb{\phi}_\ell} f = 2\,\Im\big\{\mathrm{diag}(\pmb{\gamma}_\ell^\ast) \, \bm r_\ell\big\}$
    \State Update
    $\pmb{\phi}_\ell \leftarrow \pmb{\phi}_\ell - \mu_\ell \nabla_{\pmb{\phi}_\ell} f$
    \State Project onto the unit circle
    $\pmb{\gamma}_\ell \leftarrow \exp\!\big(j\pmb{\phi}_\ell\big)$        \ElsIf{$\ell\in\mathcal{L}^{(s)}_{\mathrm{ac}}$}
    \State Compute
    $\nabla_{\pmb{\alpha}_\ell} f = 2\,\Re \big\{\mathrm{diag}(\pmb{\gamma}_\ell^\ast)\,
    \bm r_\ell\big\}$
    \State Update
    $\pmb{\alpha}_\ell \leftarrow \mathcal{P}_{[\alpha_{\min},\alpha_{\max}]}\!\left(\pmb{\alpha}_\ell-\mu_{\alpha,\ell}\nabla_{\pmb{\alpha}_\ell} f\right)$
    \State Set $\pmb{\gamma}_\ell \leftarrow \pmb{\alpha}_\ell \odot \exp\!\big(j\pmb{\phi}_\ell\big)$
    \EndIf
    \State Update
    $\bm G_0 \leftarrow \bm\Gamma_{\mathrm{dal}}\bm W_L \bm\Gamma_{L-1}\bm W_{L-1}\cdots \bm\Gamma_2\bm W_2$
    \EndFor
    \EndFor
\Statex \textit{\textbf{Step 2: ST randomization of the initial DAL (performed once per time-slot)}}
\vspace{0.2em}
\For{$m=0,1,\ldots,M-1$}
    \State Draw i.i.d. random phases $\psi_z[m]$
    \State Set: $\delta_z[m]\triangleq e^{j\psi_z[m]}$ 
    \State Set: $\pmb{\Delta}[m]\triangleq \mathrm{diag}\big(\delta_0[m],\ldots,\delta_{Z-1}[m]\big)$
    \State Compute: $\bm G[m]\triangleq \bm G_0\,\bm\Delta[m]$
\EndFor   
\end{algorithmic}
\end{algorithm}

\section{Numerical results}
\label{sec:simul}

In this section, we present Monte Carlo simulations to validate the proposed ST-SIM architecture 
and evaluate its achievable sum-rate in a multiuser downlink scenario. We first analyze an S-only (time-invariant) 
SIM configuration to quantify the impact of DAL on the synthesis accuracy, convergence behavior, and 
power efficiency of the multilayer structure through the objective function in~\eqref{eq:obj}. We then consider 
the full randomized ST-SIM framework with partial CSIT and compare it with a full-CSIT beamforming benchmark, 
evaluating the time-averaged sum-rate and fairness as functions of the user population and the number of time slots.

\begin{figure}[t]
\centering
\includegraphics[width=\linewidth]{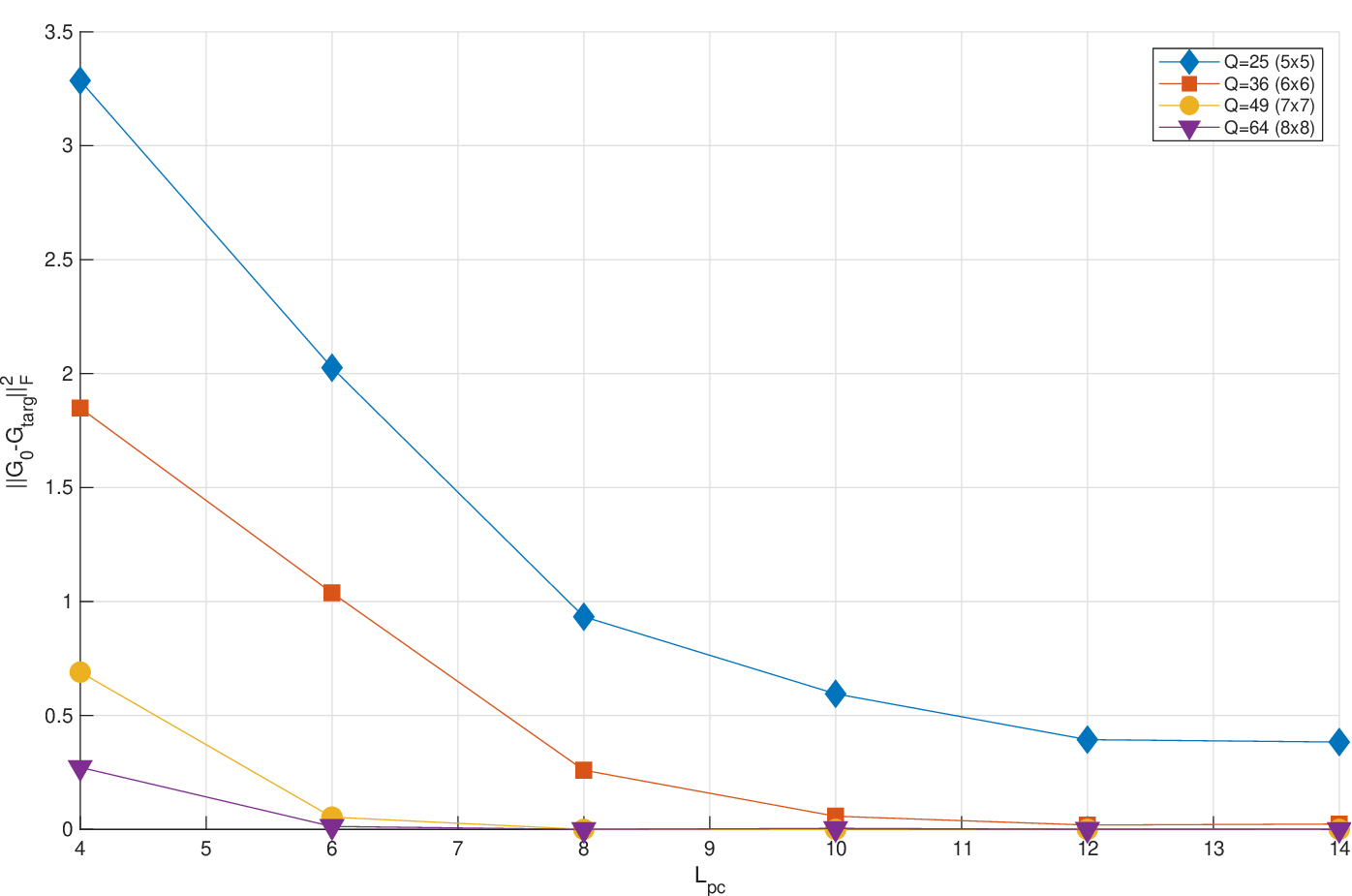}
\caption{Objective function \eqref{eq:obj}  versus the number of passive layers \( L_{\text{pc}} \) for \( Q \in \{ 25, 36 , 49 , 64\} \) meta-atoms. Parameters are set to \( Z = 9 \), \( V = 25 \), and \( L_{\text{ac}} = 4 \). 
The case \(Q=V=25\) corresponds to a baseline SIM 
architecture without absorbing meta-atoms.
All metasurface layers are square. The signal from the RF chain first passes through the AC layers and, then, through the PC layers.}
\label{fig:fig_3}
\end{figure}

\subsection{DAL-aided vs baseline S-only SIM synthesis}
\label{subsec:sim_TI_DAL}
To assess the effectiveness of incorporating the DAL, which introduces $Q - V$ perfectly absorbing elements, 
we compare the proposed SIM architecture with a conventional configuration where $Q = V$ and, thus, 
all meta-atoms are transmitting.
To this end, we consider an S-only SIM configuration whose response reduces to \eqref{eq:G0}.
In this setting, we optimize~\eqref{eq:obj}, where the target matrix $\bm{G}_{\text{targ}} \in \mathbb{C}^{V \times Z}$, 
representing the desired block response, is constructed as a Haar matrix
\be
\bm{G}_{\text{targ}} = \frac{\sqrt{N}}
{\beta \, \|\bm W_1\|} \, \bm{R}^\herm_{\text{a}} \left( \bm{R}_{\text{a}} \, \bm{R}^\herm_{\text{a}} \right)^{-1/2}
\ee
where $\bm{R}_{\text{a}} \in \mathbb{C}^{Z \times V}$ is a random matrix whose entries are independently drawn from a 
circularly symmetric complex Gaussian distribution with zero mean and unit variance.
By construction, the columns of \( \bm{G}_{\text{targ}} \) fulfill \eqref{eq:norm-constr}.
This choice prevents favoring particular spatial directions and allows assessing the synthesis capability of 
the proposed architecture in a general setting.
In this setup, we set \(Z_x=Z_y=3\) (i.e., $Z=9$), 
\(V_x=V_y=5\) (i.e., $V=25$), and   \(L_{\text{ac}}=4\).
The AC layers are placed at the input of the SIM stack, so the signal from the RF chain first passes through the AC layers and then through the PC layers. 
The placement of AC/PC layers, indeed, affects the convergence rate of the PDG algorithm, which achieves its fastest convergence when the AC layers are located before the PC ones \cite{Dar.2025}.  
\begin{figure}[t]
\centering
\includegraphics[width=\linewidth]{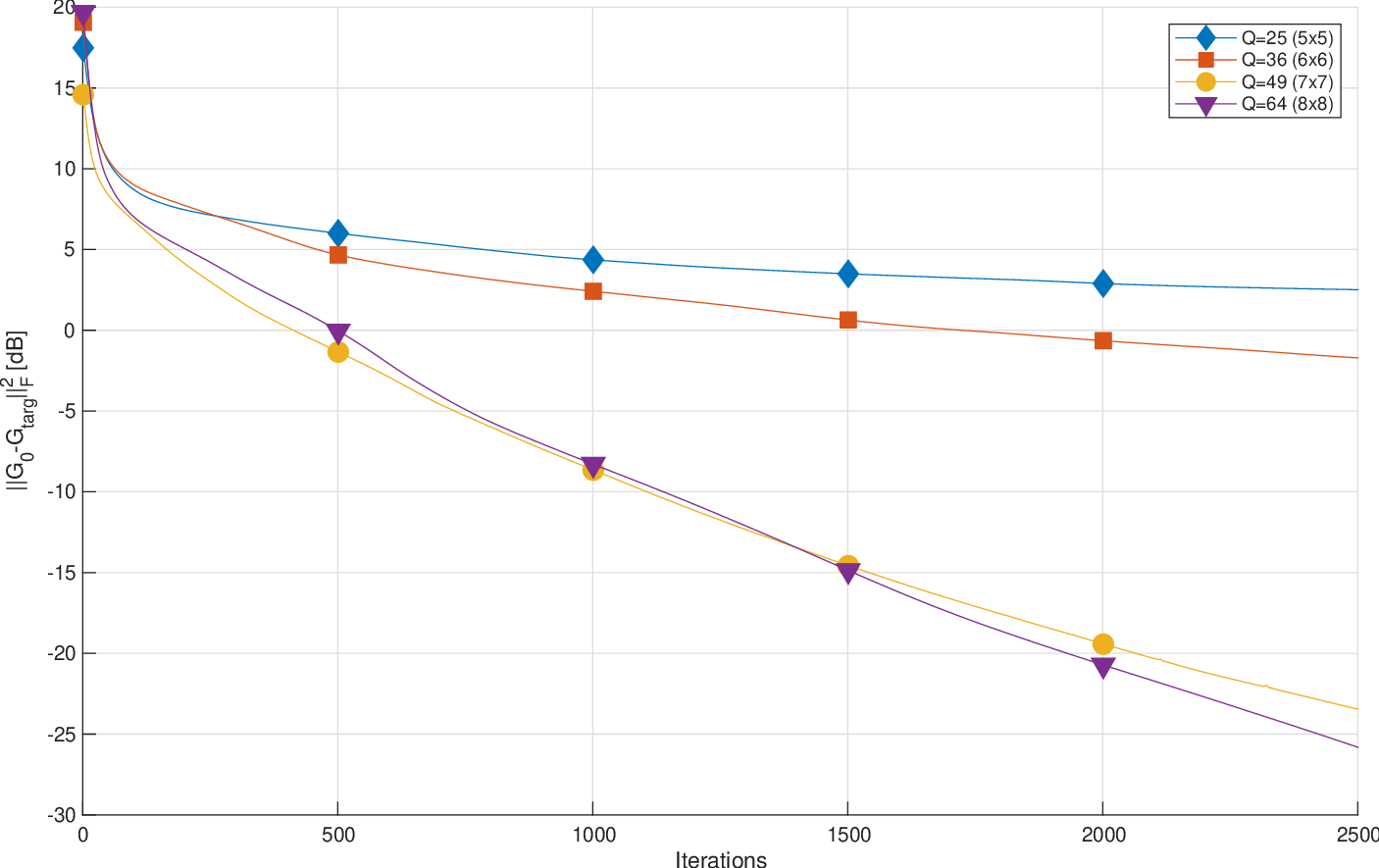}
\caption{Convergence rate of the PGD algorithm  for \( Q \in \{ 25, 36 , 49 , 64\} \) meta-atoms. Parameters are set to \( Z = 9 \), \( V = 25 \), \( L_{\text{ac}} = 4 \), and \( L_{\text{pc}} = 8 \). 
The case \(Q=V=25\) corresponds to the baseline SIM without DAL.
All metasurface layers are square. The signal from the RF chain first passes through the AC layers and, then, through the PC layers.}
\label{fig:fig_4}
\end{figure}
The amplitude responses of the AC layers satisfy \( \alpha_{\ell,q} \in [\alpha_{\text{min}}, \alpha_{\text{max}}] \), 
with \( \alpha_{\text{min}} = -22 \) dB and \( \alpha_{\text{max}} = 13 \) dB \cite{Liu.2022}, for all \( \ell \in \mathcal{L}_{\text{ac}}^{\text{(s)}} \) and \( q \in \mathcal{Q} \).

Figure~\ref{fig:fig_3} reports the objective function~\eqref{eq:obj} as a function of the number of PC layers \(L_{\mathrm{pc}}\), ranging from 4 to 14, for different values of the number of meta-atoms \(Q\). The PC layers are assumed to have constant transmittance \(\alpha_{\mathrm{pc}}=0.9\).
When \(Q=V=25\), the considered transmit structure reduces to a baseline SIM architecture of size \(Z\times Q\) with no absorbing meta-atoms. In this case, the synthesis of the target matrix \(\bm{G}_{\text{targ}}\in\mathbb{C}^{Q\times Z}\) depends only on the number of passive layers \(L_{\mathrm{pc}}\). By contrast, in the DAL-aided SIM both \(L_{\mathrm{pc}}\) and \(Q\) act as design degrees of freedom.
As shown in Fig.~\ref{fig:fig_3}, the DAL-aided SIM with \(Q-V\) absorbing meta-atoms consistently outperforms the baseline configuration for all values of \(L_{\mathrm{pc}}\). For a fixed number of meta-atoms \(Q\), the performance improves as the number of PC layers increases. Moreover, for a fixed number of passive layers \(L_{\mathrm{pc}}\), the objective function~\eqref{eq:obj} decreases as \(Q\) increases when the DAL is employed (\(Q>V\)). This additional flexibility enables arbitrarily small approximation errors \(\|\bm{G}_0-\bm{G}_{\text{targ}}\|^2\). Conversely, in the baseline SIM the objective function ceases to decrease monotonically beyond a certain number of layers due to error propagation in the PGD iterations~\cite{DiRenzo}.

Figure~\ref{fig:fig_4} shows the convergence behavior of the PGD algorithm for \(Q\in\{25,36,49,64\}\) with \(L_{\mathrm{pc}}=8\). The objective function 
\(\|\bm{G}_0-\bm{G}_{\text{targ}}\|^2\) decreases with the number of iterations for all configurations. Increasing \(Q\) significantly improves convergence, yielding both faster decay and lower steady-state error, with values below approximately \(-25\,\mathrm{dB}\). This confirms that enlarging the number of meta-atoms increases the available degrees of freedom in the DAL-aided synthesis and improves the approximation of \(\bm{G}_{\text{targ}}\). The case \(Q=25\) corresponds to the baseline SIM (without DAL) and exhibits the worst performance, whereas larger values of \(Q\) provide substantial gains due to the additional absorbing meta-atoms introduced by the DAL.

\begin{figure}[t]
\centering
\includegraphics[width=\linewidth,trim=40 250 60 265,clip]{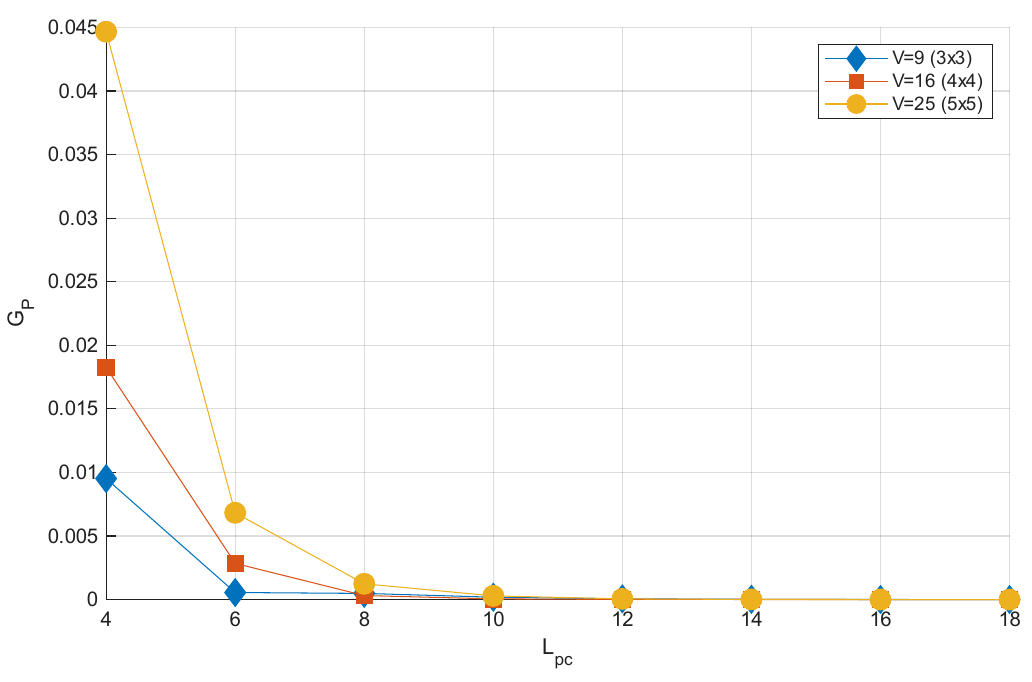}
\caption{Power gain of a DAL-aided and conventional SIM employing only PC layers  for \( V \in \{ 9, 16 , 25\} \) meta-atoms. Parameters are set to \( Z = 9 \) and \( Q = 25 \). The case \(Q=V=25\) corresponds to a baseline SIM architecture without absorbing meta-atoms. All metasurface layers are square.}
\label{fig:fig_7}
\end{figure}

We now analyze the impact of the DAL on the power efficiency of the SIM. 
The presence of absorbing meta-atoms introduces intrinsic power dissipation, since part of the incident EM energy is absorbed rather than transferred through the multilayer structure. Therefore, it is of interest to quantify how efficiently the SIM transfers power from input to output and compare it with a baseline configuration where all meta-atoms are transmitting.
In this respect, we define the \emph{power gain} of the SIM as
\begin{equation}
G_{\mathrm P} \triangleq \frac{\euscr{P}_{\mathrm{rad}}}{\euscr{P}_{\mathrm{sig}}}
\end{equation}
where $\euscr{P}_{\mathrm{rad}}$ is the radiated power in~\eqref{eq:Prad} and $\euscr{P}_{\mathrm{sig}}$ is the input signal power. 
Although the total energy efficiency also depends on the consumption of the active layers, this contribution is implementation-dependent. Hence, we focus on the technology-agnostic metric $G_{\mathrm P}$.
From~\eqref{eq:Prad}, the radiated power becomes $\euscr{P}_{\mathrm{rad}} = (\euscr{P}_{\mathrm{sig}}/N)\|\bm G_0\|^2$, 
hence yielding
\begin{equation}
G_{\mathrm P} = \frac{\|\bm G_0\|^2}{N} \:.
\end{equation}
A purely passive phase-only SIM generally exhibits a gain significantly below the bound implied by~\eqref{eq:norm-constr} due to propagation losses, whereas the inclusion of amplitude-controlled (active) layers allows approaching this limit through local amplification.
The analytical evaluation of $G_{\mathrm P}$ is complicated by the cascaded products of diagonal transmission matrices and coupling matrices. Therefore, we rely on numerical results to compare the DAL-aided and baseline architectures.

\begin{figure}[t] 
\centering 
\includegraphics[width=\linewidth,trim=40 250 60 265,clip]{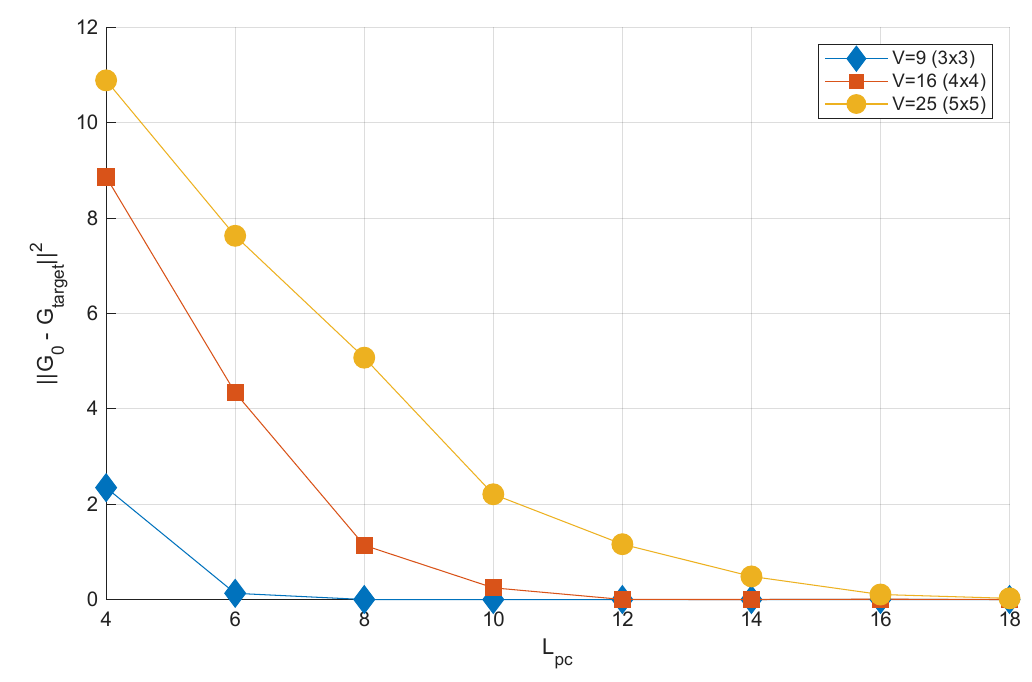} \caption{Objective function \eqref{eq:obj} of a DAL-aided and conventional SIM employing only PC layers versus \( L_\text{pc} \). Parameters are set to \( Z = 9 \) and \( Q = 25 \). The case \(Q=V=25\) corresponds to a baseline SIM architecture without absorbing meta-atoms. All metasurface layers are square.} 
\label{fig:fig_8} 
\end{figure}

We consider $Z=9$, $Q=25$, and $V\in\{9,16,25\}$, where $V=Q$ corresponds to the baseline SIM. 
Figure~\ref{fig:fig_7} reports the power gain for a purely passive configuration employing only PC layers as a function of $L_{\mathrm{pc}}$.
The baseline SIM achieves a higher power gain than the DAL-aided configurations, and the gap increases as the number of absorbing meta-atoms grows (i.e., decreasing $V$). However, this comparison does not account for synthesis accuracy. As shown in Fig.~\ref{fig:fig_8}, the baseline SIM reaches a squared error comparable to the DAL-aided architecture only for $L_{\mathrm{pc}}\geq16$, whereas the DAL-aided SIM attains low error with significantly fewer layers. Hence, in passive implementations the DAL trades a moderate reduction in power gain for a substantially lower number of layers required for convergence.

\begin{figure}[t]
\centering
\includegraphics[width=\linewidth,trim=40 250 60 265,clip]{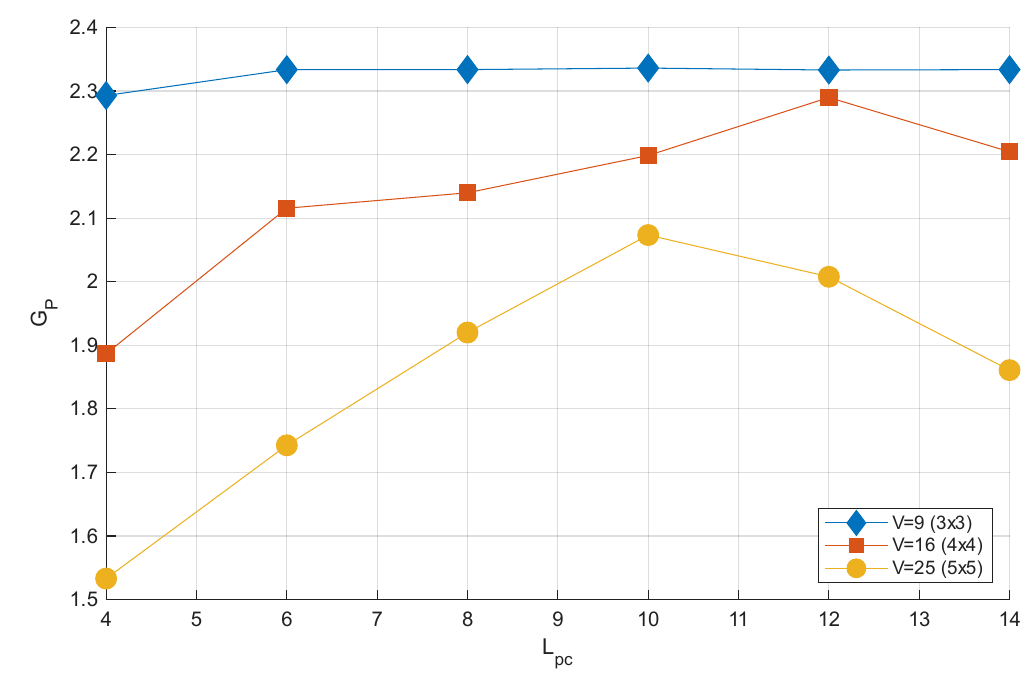}
\caption{Power gain of a DAL-aided and conventional SIM employing PC and AC layers  for \( V \in \{ 9, 16 , 25\} \) meta-atoms. Parameters are set to \( Z = 9 \), \( Q = 25 \), and $L_\text{ac}=4$. The case \(Q=V=25\) corresponds to a baseline SIM architecture without absorbing meta-atoms.
All metasurface layers are square. The signal from the RF chain first passes through the AC layers and, then, through the PC layers.} 
\label{fig:fig_9}
\end{figure}

\begin{figure}[t]
\centering
\includegraphics[width=\linewidth,trim=40 250 60 265,clip]{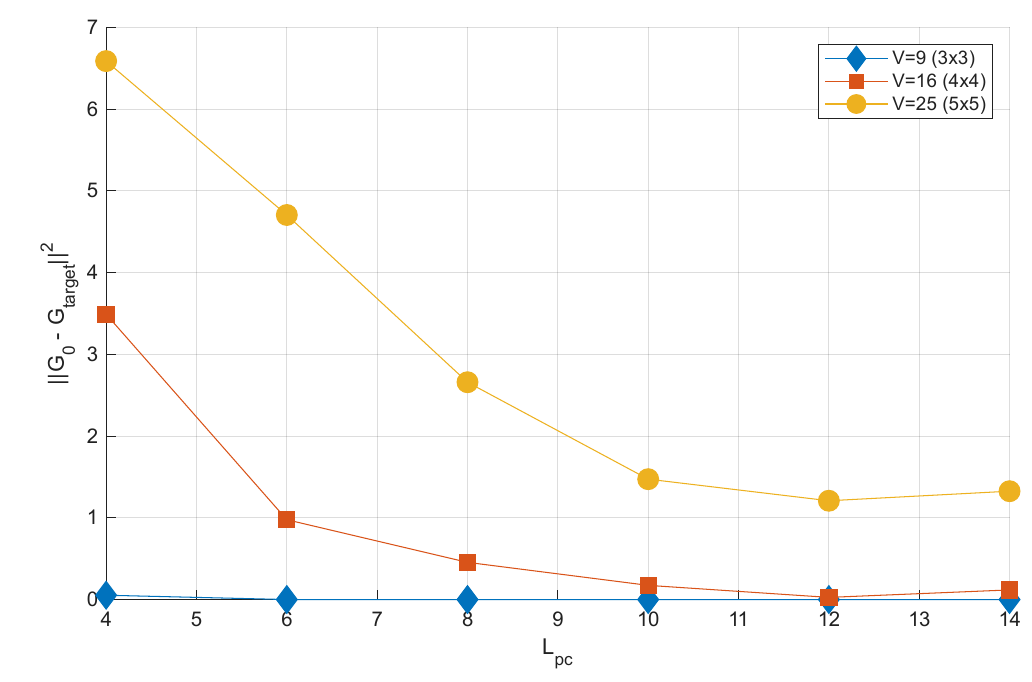}
\caption{Objective function \eqref{eq:obj} of a DAL-aided and conventional SIM employing PC and AC layers  versus \( L_\text{pc} \) for \( V \in \{ 9, 16 , 25\} \) meta-atoms. Parameters are set to \( Z = 9 \), \( Q = 25 \), and $L_\text{ac}=4$. The case \(Q=V=25\) corresponds to a baseline SIM architecture without absorbing meta-atoms.
All metasurface layers are square. The signal from the RF chain first passes through the AC layers and, then, through the PC layers.} 
\label{fig:fig_10}
\end{figure}

Figure~\ref{fig:fig_9} shows the power gain when both passive and active layers are used ($L_{\mathrm{ac}}=4$). 
In this case, the DAL-aided architectures ($V<Q$) outperform the baseline SIM for all values of $L_{\mathrm{pc}}$, with the best performance obtained for $V=9$. 
This behavior indicates that, in the presence of active layers, the additional degrees of freedom introduced by the DAL improve not only the approximation accuracy but also the capability of concentrating power on the transmitting meta-atoms. This observation is consistent with the objective-function behavior shown in Fig.~\ref{fig:fig_10}.

\subsection{Multiuser downlink performance of the ST-SIM}
\label{subsec:sim_beamforming}
Herein, we evaluate the effective sum-rate achieved by the proposed randomized ST-SIM multiuser downlink. 
The BS is located at $(0,0,h_{\mathrm{BS}})$ with $h_{\mathrm{BS}}=10$ m. The $U$ users are uniformly distributed in a circular annulus centered at the origin on the $xy$-plane, with inner radius $r_i=10$ m and outer radius $r_o=100$ m. The channel variance toward the $u$-th user is $\sigma_u^2 = d_u^{-\eta}\lambda_0^2/(4\pi)^2$, where $d_u$ denotes the link distance and $\eta=1.6$. The coherence time is $T=5$ ms.
The carrier frequency is $f_0=28$ GHz with bandwidth $10$ MHz, roll-off factor $\rho_s=0.25$, and noise spectral density $-174$ dBm/Hz. The total radiated power is $\euscr{P}_{\mathrm{rad}}=15$ dBm.
The BS employs a UPA with spacing $d_{\mathrm{meta}}=\lambda_0/2$ and $N_x=N_y=2$ ($N=4$). The ST-SIM layers are spaced by $s_{\mathrm{lay}}=\lambda_0/2$. The initial DAL uses $Z_x=Z_y=3$ meta-atoms, intermediate layers use $Q_x=Q_y=24$, and the number of PC and AC layers is $L_{\mathrm{pc}}=6$ and $L_{\mathrm{ac}}=2$, respectively.

As benchmark, we consider a fully-digital MIMO system with full CSIT, where each user feeds back its channel vector $\bm h_u$. 
The transmitter selects $N$ users and applies linear precoding using normalized channel directions 
$\widetilde{\bm g}_{u_n^\star}=\bm h_{u_n^\star}/\|\bm h_{u_n^\star}\|^2$ \cite{Tse-book}. 
This scheme represents a best-case reference in terms of beamforming capability under full CSIT. In contrast, the 
proposed ST-SIM architecture operates with only $N$ RF chains while exploiting a larger metasurface aperture 
with $V$ meta-atoms for wave-domain beamforming.
For the proposed ST-SIM architecture, the overhead factor in \eqref{eq:Reff} is 
\begin{equation}
\xi_{\mathrm{ST\text{-}SIM}} = 1-\frac{M(N+1)}{L_c}
\end{equation}
whereas for the conventional MIMO benchmark we use
\begin{equation}
\xi_{\mathrm{MIMO}} = 1-\frac{2V}{L_c}
\end{equation}
where $L_c$ denotes the number of symbol periods in each channel coherence interval.

\begin{figure}[t] 
\centering 
\includegraphics[width=\linewidth]{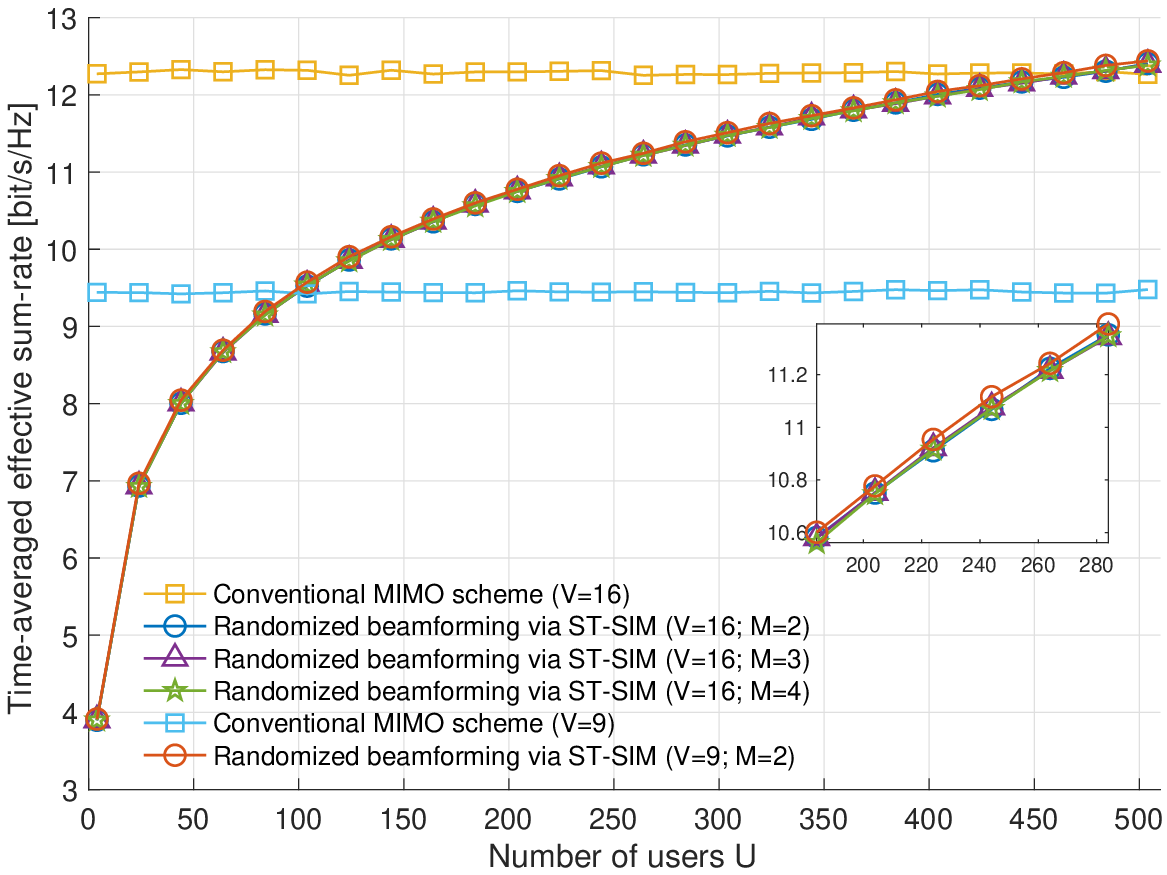} 
\caption{TA effective sum-rate as a function of the number of users $U$ 
for three values of the number of time-slots $M$, with both 
$V=9$ and $V=16$ meta-atoms. Other ST-SIM parameters are 
\( Q = 576 \), \( Z = 9 \), \( N = 4 \), \( L_{\text{ac}} = 2 \), and \( L_{\text{pc}} = 6 \). All metasurface layers are square. The signal from the RF chain first passes through the AC layers and, then, through the PC layers.} 
\label{fig:fig_5} 
\end{figure}

Figure~\ref{fig:fig_5} shows the time-averaged effective sum-rate versus the number of users $U$. 
We consider $V\in\{9,16\}$ and set $M$ according to~\eqref{eq:rel}. 
The PGD algorithm uses $\kappa_{\max}=10^3$, which ensures that 
$\|\bm G_0-\bm G_{\mathrm{targ}}\|^2=10^{-8}$ for $V=9$ and $10^{-5}$ for $V=16$.
As $U$ increases, the effective sum-rate of the proposed scheme grows for all values of $M$ and $V$. 
For $V=9$, the randomized ST-SIM outperforms the full-CSIT MIMO scheme for $U>100$, while for $V=16$ the crossover occurs around $U>500$. 
The gain arises from multiuser diversity enabled by channel randomization, whereas the digital benchmark benefits from perfect amplitude and phase knowledge at the transmitter at the cost of significantly higher feedback (see Remark~3).

\begin{figure}[t]
\centering
\includegraphics[width=1.1\linewidth]{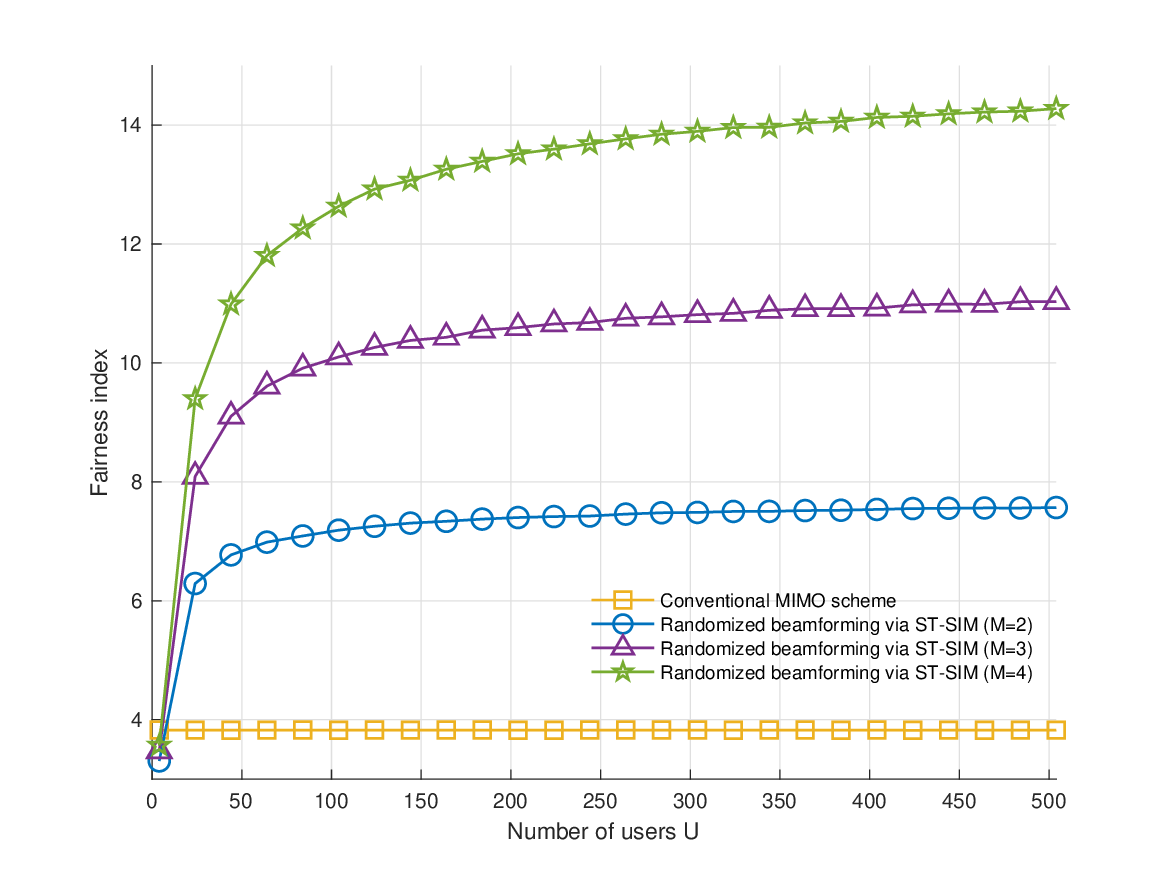}
\caption{Fairness index as a function of the number of users $U$ for 
three values of the number of time-slots $M$. 
ST-SIM parameters are set to \( Q = 576 \), \( Z = 100 \), \( V = 9 \), \( N = 4 \), \( L_{\text{ac}} = 2 \), and \( L_{\text{pc}} = 6 \). All metasurface layers are square. The signal from the RF chain first passes through the AC layers and, then, through the PC layers.}
\label{fig:fig_6}
\end{figure}

In order to assess the fairness gain of the randomized ST SIM-based beamforming relative to the conventional MIMO scheme, we adopt the following fairness index \cite{Jain.1984}, 
averaged over $M$ time slots
\be
\euscr{F} \eqdef \frac{1}{M}\sum_{m=0}^{M-1}  \frac{[\sum_{u=1}^U R_u(m)]^2}{\sum_{u=1}^U R^2_u(m)} 
\ee
which is presented in Fig.~\ref{fig:fig_6} as a function of the number of users $U$ for three different values of $M$.
Results show that in the conventional MIMO scheme only $N=4$ users are scheduled on average for transmission, whereas the proposed randomized beamforming increases by a factor $M$ the average number of scheduled users thanks to its randomization-induced multiuser diversity. 
Larger values of $M$ enhance multiuser diversity but also increase training and CSI-feedback overheads (see Remark~2).

\section{Conclusions}
\label{sec:concl}

This paper introduced a beamforming framework for massive multiuser downlink connectivity based on randomized 
ST-SIM. By combining a time-varying DAL with multiple S-only metasurface layers, the architecture enables 
joint spatial–temporal wavefront shaping, increasing scheduling flexibility and exploiting multiuser 
diversity even under slowly varying channel conditions.
A signal and propagation model was developed together with a PGD synthesis algorithm 
for configuring the metasurface coefficients. To reduce channel acquisition and feedback overhead, a 
partial-CSIT beamforming strategy based on randomized steering vectors and low-rate user feedback was proposed.
Numerical results demonstrated that the randomized ST-SIM achieves competitive sum-rate performance with 
significantly reduced signaling requirements compared to full-CSIT solutions, supporting scalable 
downlink operation in dense networks.

\begin{IEEEbiography}[{\includegraphics[width=1in,height=1.25in,clip,keepaspectratio]
{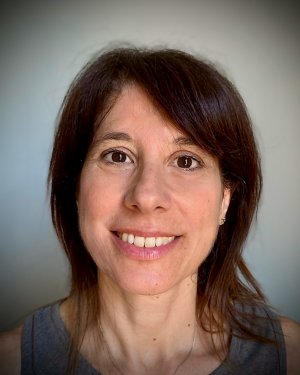}}]
{Donatella Darsena} (Senior Member, IEEE) received the Dr. Eng. degree summa cum laude in telecommunications engineering in 2001, and the Ph.D. degree in electronic and telecommunications engineering in 2005, both from the University of Napoli Federico II, Italy. From 2001 to 2002, she worked as embedded system designer in the Telecommunications, Peripherals and Automotive Group, STMicroelectronics, Milano, Italy.
In 2005 she joined the Department of Engineering at Parthenope University of Napoli, Italy and worked first as an Assistant Professor and then as an Associate Professor from 2005 to 2022.
She is currently an Associate Professor in the Department of Electrical Engineering and Information Technology of the University of Napoli Federico II, Italy.
Her research interests are in the broad area of signal processing for communications, with current emphasis on reflected-power communications, orthogonal and nonorthogonal multiple access techniques, wireless system optimization, and physical-layer security.
Dr. Darsena has served as a Senior Editor for IEEE SIGNAL PROCESSING LETTERS since 2026, Senior Editor for IEEE ACCESS since 2024, and Executive Editor for IEEE COMMUNICATIONS LETTERS since 2023. She was an Associate Editor of IEEE ACCESS (from 2018 to 2023), of IEEE SIGNAL PROCESSING LETTERS (from 2020 to 2025), and Senior Area Editor of IEEE COMMUNICATIONS LETTERS (from 2020 to 2023).
\end{IEEEbiography}

\begin{IEEEbiography}
[{\includegraphics[width=1in,height=1.25in,clip,keepaspectratio]
{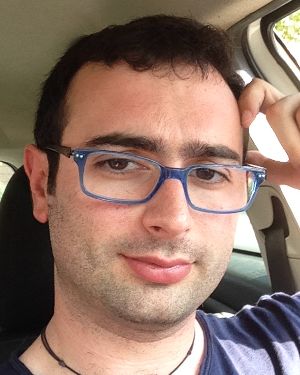}}]
{Ivan Iudice} (Senior Member, IEEE) was born in Livorno, Italy, in 1986. He received the B.S. and M.S. degrees in telecommunications engineering in 2008 and 2010, respectively, and the Ph.D. degree in information technology and electrical engineering in 2017, all from University of Napoli Federico II, Italy. Since 2011, he has been with the Italian Aerospace Research Centre (CIRA), Capua, Italy. He first served as part of the Electronics and Communications Laboratory and he is currently part of the Security and Cybersecurity in Complex Systems and Air Traffic Management Unit. His research activities mainly lie in the area of signal and array processing for communications, with current interests focused on physical-layer security, space-time techniques for cooperative communications systems and reconfigurable metasurfaces. He is involved in several international projects. Since 2025 he has been secretary of the IEEE AESS technical panel: “Glue Technologies for Space Systems. He serves as reviewer for several international journals and as TPC member for several international conferences. He is author of several papers on refereed journals and international conferences. 
He has been served as a Associate Editor for IEEE SIGNAL PROCESSING LETTERS since 2025.
\end{IEEEbiography}

\begin{IEEEbiography}[
{\includegraphics[width=1in,height=1.25in,clip,keepaspectratio]
{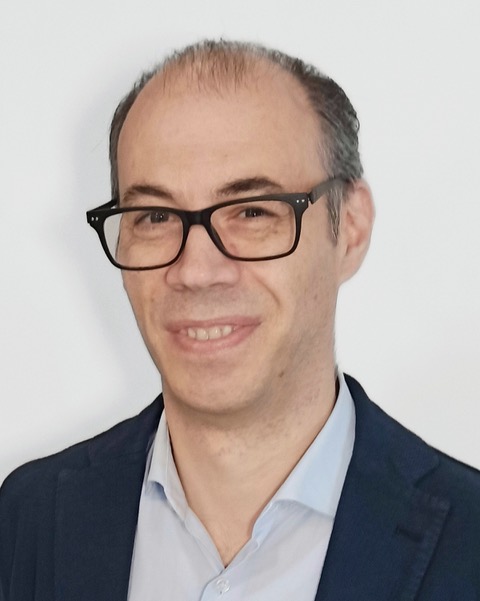}}]
{Vincenzo Galdi} (Fellow, IEEE) received the Laurea degree (\textit{summa cum laude}) in electrical engineering and the Ph.D. degree in applied electromagnetics from the University of Salerno, Salerno, Italy, in 1995 and 1999, respectively. He held several research and visiting appointments at international institutions, including the European Space Research and Technology Centre (ESTEC), Noordwijk, The Netherlands; Boston University, Boston, MA, USA; the Massachusetts Institute of Technology, Cambridge, MA, USA; the California Institute of Technology, Pasadena, CA, USA; and The University of Texas at Austin, Austin, TX, USA. He is currently a Professor of Electromagnetics with the Department of Engineering, University of Sannio, Benevento, Italy, where he leads the Fields \& Waves Lab. He is also the co-founder of the spinoff company MANTID, Benevento, Italy, and the startup company BioTag, Naples, Italy. He has co-edited two books, coauthored over 200 papers in peer-reviewed international journals, and is a co-inventor of 13 patents. His research interests include wave interactions with complex structures and media, multiphysics metamaterials, smart propagation environments, optical sensing, and gravitational interferometry. Dr.~Galdi received the URSI Young Scientist Award in 2001 and the Outstanding Associate Editor Award of the \textit{IEEE Transactions on Antennas and Propagation} in 2014. He served as Chair of the Technical Program Committee of the International Congress on Engineered Material Platforms for Novel Wave Phenomena in 2018. He also served as a Topical/Track Chair for the \textit{IEEE International Symposium on Antennas and Propagation} and the USNC-URSI Radio Science Meeting (2016--2017 and 2020--2023), and as Organizer/Chair of several topical workshops and special sessions. He served as an Associate Editor (2013--2014), Senior Associate Editor (2015--2016), and Track Editor (2016--2020) of the \textit{IEEE Transactions on Antennas and Propagation}. He also served as an Associate Editor of \textit{Optics Express} (2019--2025) and as a regular reviewer for numerous journals, conferences, and funding agencies. He is currently serving as an AdCom Member (IEEE Antennas and Propagation Society representative) of the IEEE Nanotechnology Council. Dr.~Galdi is a Fellow of Optica and a Fellow of the American Physical Society. He is also a Senior Member of the LIGO Scientific Collaboration. 
\end{IEEEbiography}

\begin{IEEEbiography}[
{\includegraphics[width=1in,height=1.25in,clip,keepaspectratio]
{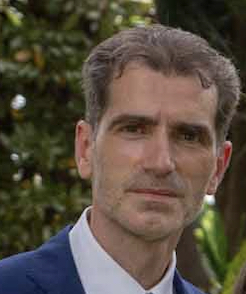}}]
{Francesco Verde} (Senior Member, IEEE) received the Dr. Eng. degree (summa cum laude) in electronic engineering from the Second University of Naples (SUN), Aversa, Italy, in 1998, and the Ph.D. degree in information engineering from the University of Naples Federico II, Italy, in 2002.
From December 2002 to August 2025, he was with the University of Naples Federico II, where he served as an Assistant Professor of signal theory and mobile communications and was promoted to Associate Professor of telecommunications in December 2011. Since September 2025, he has been an Associate Professor of telecommunications with the Department of Engineering, University of Campania Luigi Vanvitelli, Aversa, Italy.

His research interests include wireless communications via metamaterials, reflected-power communications, physical-layer security, unmanned aerial vehicle (UAV)-assisted communications, orthogonal and non-orthogonal multiple-access techniques, and space–time processing for cooperative and cognitive communication systems.
Prof. Verde has served on numerous technical program committees for major IEEE conferences in signal processing and wireless communications. He has been an Associate Editor of the IEEE Transactions on Vehicular Technology since 2022 and the IEEE Transactions on Wireless Communications since 2026. He was previously an Associate Editor of the IEEE Transactions on Signal Processing (2010–2014), IEEE Signal Processing Letters (2014–2018), and IEEE Transactions on Communications (2017–2022), as well as a Senior Area Editor of IEEE Signal Processing Letters (2018–2023). He also served as a Guest Editor for the EURASIP Journal on Advances in Signal Processing (2010) and Sensors (MDPI) (2018–2022).
\end{IEEEbiography}


\begin{thebibliography}{99}

\bibitem{ITU.2023}
\emph{Framework and Overall Objectives of the Future Development of IMT for 2030 and Beyond}, 
ITU-R M.2160-0, Nov.~2023.

\bibitem{Kalor.2024}
A.~E.~Kalor \emph{et al.}, 
``Wireless 6G connectivity for massive number of devices and critical services,'' 
\emph{Proc. IEEE}, vol.~113, pp.~826--848, Sep.~2025.

\bibitem{Hanzo}
J.~An \emph{et al.}, 
``Stacked intelligent metasurfaces for efficient holographic MIMO communications in 6G,'' 
\IeeeJSAC, vol.~41, pp.~2380--2396, Aug.~2023.

\bibitem{Lin.2018}
X.~Lin, Y.~Rivenson, N.~T.~Yardimci \emph{et al.}, 
``All-optical machine learning using diffractive deep neural networks,'' 
\emph{Science}, vol.~361, pp.~1004--1008, 2018.

\bibitem{Liu.2022}
C.~Liu \emph{et al.}, 
``A programmable diffractive deep neural network based on a digital-coding metasurface array,'' 
\emph{Nature Electron.}, vol.~5, pp.~113--122, Feb.~2022.

\bibitem{Basar.2024}
E.~Basar \emph{et al.}, 
``Reconfigurable intelligent surfaces for 6G: Emerging hardware architectures, applications, and open challenges,'' 
\emph{IEEE Veh. Technol. Mag.}, vol.~19, pp.~27--47, Sep.~2024.

\bibitem{Hassan.2024}
N.~U.~Hassan, J.~An, M.~Di~Renzo, M.~Debbah, and C.~Yuen, 
``Efficient beamforming and radiation pattern control using stacked intelligent metasurfaces,'' 
\emph{IEEE Open J. Commun. Soc.}, vol.~5, pp.~599--611, 2024.

\bibitem{Nerini.2024}
M.~Nerini and B.~Clerckx, 
``Physically consistent modeling of stacked intelligent metasurfaces implemented with beyond diagonal RIS,'' 
\IeeeWCOMMLETT, vol.~28, pp.~1693--1697, Jul.~2024.

\bibitem{DiRenzo}
J.~An, C.~Yuen, Y.~L.~Guan, M.~Di~Renzo, M.~Debbah, H.~V.~Poor, and L.~Hanzo, 
``Two-dimensional direction-of-arrival estimation using stacked intelligent metasurfaces,'' 
\IeeeJSAC, vol.~42, pp.~2786--2802, Oct.~2024.

\bibitem{Yao.2024}
X.~Yao, J.~An, L.~Gan, M.~Di~Renzo, and C.~Yuen, 
``Channel estimation for stacked intelligent metasurface-assisted wireless networks,'' 
\IeeeWCOMMLETT, vol.~13, pp.~1349--1353, May~2024.

\bibitem{Pap.2025}
A.~Papazafeiropoulos, P.~Kourtessis, S.~Chatzinotas, D.~I.~Kaklamani, and I.~S.~Venieris, 
``Performance of double-stacked intelligent metasurface-assisted multiuser massive MIMO communications in the wave domain,'' 
\IeeeTWC, vol.~24, pp.~4205--4218, May~2025.

\bibitem{DiRenzo-ICC}
J.~An, M.~Di~Renzo, M.~Debbah, and C.~Yuen, 
``Stacked intelligent metasurfaces for multiuser beamforming in the wave domain,'' 
in \emph{Proc. IEEE Int. Conf. Commun. (ICC)}, Rome, Italy, May~2023, pp.~2834--2839.

\bibitem{Liu.2024}
H.~Liu, J.~An, D.~W.~Kwan Ng, G.~C.~Alexandropoulos, and L.~Gan, 
``DRL-based orchestration of multi-user MISO systems with stacked intelligent metasurfaces,'' 
in \emph{Proc. IEEE Int. Conf. Commun. (ICC)}, Denver, CO, USA, 2024, pp.~4991--4996.

\bibitem{An_ArXiv_2025}
J.~An, M.~Di~Renzo, M.~Debbah, H.~V.~Poor, and C.~Yuen, 
``Stacked intelligent metasurfaces for multiuser downlink beamforming in the wave domain,'' 
\IeeeTWC, vol.~24, pp.~5525--5538, Jul.~2025.

\bibitem{Lin.2024}
S.~Lin, J.~An, L.~Gan, M.~Debbah, and C.~Yuen, 
``Stacked intelligent metasurface enabled LEO satellite communications relying on statistical CSI,'' 
\IeeeWCOMMLETT, vol.~13, pp.~1295--1299, May~2024.

\bibitem{Pap.2024}
A.~Papazafeiropoulos, P.~Kourtessis, S.~Chatzinotas, D.~I.~Kaklamani, and I.~S.~Venieris, 
``Achievable rate optimization for large stacked intelligent metasurfaces based on statistical CSI,'' 
\IeeeWCOMMLETT, vol.~13, pp.~2337--2341, Sep.~2024.

\bibitem{Li.2025}
Q.~Li, M.~El-Hajjar, C.~Xu, J.~An, C.~Yuen, and L.~Hanzo, 
``Stacked intelligent metasurface-based transceiver design for near-field wideband systems,'' 
\IeeeTCOMM, vol.~73, pp.~8125--8139, Sep.~2025.

\bibitem{Dar.2025}
D.~Darsena, F.~Verde, I.~Iudice, and V.~Galdi, 
``Design of stacked intelligent metasurfaces with reconfigurable amplitude and phase for multiuser downlink beamforming,'' 
\emph{IEEE Open J. Commun. Soc.}, vol.~6, pp.~531--550, 2025.

\bibitem{Li_2024}
Q.~Li, M.~El-Hajjar, C.~Xu, J.~An, C.~Yuen, and L.~Hanzo, 
``Stacked intelligent metasurfaces for holographic MIMO-aided cell-free networks,'' 
\IeeeTCOMM, vol.~72, pp.~7139--7151, Nov.~2024.

\bibitem{Park-arXiv_2025}
E.~Park, S.-H.~Park, O.~Simeone, M.~Di~Renzo, and S.~Shamai, 
``SIM-enabled hybrid digital-wave beamforming for fronthaul-constrained cell-free massive MIMO systems,'' 
\IeeeTWC, vol.~25, pp.~7014--7031, 2026.

\bibitem{Hu_2025}
Y.~Hu \emph{et al.}, 
``Joint beamforming and power allocation design for stacked intelligent metasurfaces-aided cell-free massive MIMO systems,'' 
\IeeeTVT, vol.~74, pp.~5235--5240, Mar.~2025.

\bibitem{Shi_2025-May}
E.~Shi, J.~Zhang, Y.~Zhu, J.~An, C.~Yuen, and B.~Ai, 
``Uplink performance of stacked intelligent metasurface-enhanced cell-free massive MIMO systems,'' 
\IeeeTWC, vol.~24, pp.~3731--3746, May~2025.

\bibitem{Shi_2025-June}
E.~Shi \emph{et al.}, 
``Joint AP-UE association and precoding for SIM-aided cell-free massive MIMO systems,'' 
\IeeeTWC, vol.~24, pp.~5352--5367, Jun.~2025.

\bibitem{Shi-2025}
E.~Shi, J.~Zhang, J.~An, M.~Di~Renzo, B.~Ai, and C.~Yuen, 
``Energy-efficient SIM-assisted communications: How many layers do we need?,'' 
\IeeeTWC, Early Access, 2025.

\bibitem{Li-2024}
Q.~Li, M.~El-Hajjar, C.~Xu, J.~An, C.~Yuen, and L.~Hanzo, 
``Stacked intelligent metasurfaces for holographic MIMO-aided cell-free networks,'' 
\IeeeTCOMM, vol.~72, pp.~7139--7151, Nov.~2024.

\bibitem{Li-2025}
Q.~Li, M.~El-Hajjar, K.~Cao, C.~Xu, H.~Haas, and L.~Hanzo, 
``Holographic metasurface-based beamforming for multi-altitude LEO satellite networks,'' 
\IeeeTWC, vol.~24, pp.~3103--3116, Apr.~2025.

\bibitem{Li-2024-hw}
Q.~Li, M.~El-Hajjar, Y.~Sun, and L.~Hanzo, 
``Performance analysis of reconfigurable holographic surfaces in the near-field scenario of cell-free networks under hardware impairments,'' 
\IeeeTWC, vol.~23, pp.~11972--11984, Sep.~2024.

\bibitem{Zhang.2018}
L.~Zhang \emph{et al.}, 
``Space-time-coding digital metasurfaces,'' 
\emph{Nature Commun.}, vol.~9, Art.~no.~4334, 2018.

\bibitem{Minkov_2017}
M.~Minkov, Y.~Shi, and S.~Fan, 
``Exact solution to the steady-state dynamics of a periodically modulated resonator,'' 
\emph{APL Photon.}, vol.~2, no.~7, Art.~no.~076101, Jul.~2017.

\bibitem{Goodman}
J.~W.~Goodman, 
\emph{Introduction to Fourier Optics}, 4th~ed. 
New York, NY, USA: McGraw-Hill, 2017.

\bibitem{Tsitsas.2017}
N.~L.~Tsitsas and C.~A.~Valagiannopoulos, 
``Anomalous reflection of visible light by all-dielectric gradient metasurfaces,'' 
\emph{J. Opt. Soc. Am. B}, vol.~34, no.~7, pp.~D1--D8, Jul.~2017.

\bibitem{Wang.2023}
T.~Wang, J.~Han, X.~Ma, H.~Liu, and L.~Li, 
``Frequency-diverse MIMO metasurface antenna for computational imaging with aperture rotation technique,'' 
\emph{Frontiers Mater.}, vol.~9, Art.~no.~1112339, Jan.~2023.

\bibitem{Li.2026}
Z.~Li, J.~An, and C.~Yuen, 
``Stacked intelligent metasurface-enhanced MIMO OFDM wideband communication systems,'' 
\IeeeTWC, vol.~25, pp.~9608--9622, 2026.

\bibitem{Li.2026-IM}
Z.~Li, J.~An, and C.~Yuen, 
``Stacked intelligent metasurface-enhanced wideband multiuser MIMO OFDM-IM communications,'' 
\emph{arXiv:2509.22327}, Sep.~2025.

\bibitem{Caire}
G.~Caire, N.~Jindal, M.~Kobayashi, and N.~Ravindran, 
``Multiuser MIMO achievable rates with downlink training and channel state feedback,'' 
\IeeeTIT, vol.~56, pp.~2845--2866, Jun.~2010.

\bibitem{Marzetta}
T.~L.~Marzetta and B.~M.~Hochwald, 
``Fast transfer of channel state information in wireless systems,'' 
\IeeeTSP, vol.~54, pp.~1268--1278, Apr.~2006.

\bibitem{Kay-book}
S.~M.~Kay, 
\emph{Fundamentals of Statistical Signal Processing: Estimation Theory}. 
Prentice-Hall, 1993.

\bibitem{Vis-2002}
P.~Viswanath, D.~Tse, and R.~Laroia, 
``Opportunistic beamforming using dumb antennas,'' 
\IeeeTIT, vol.~48, pp.~1277--1294, Jun.~2002.

\bibitem{Love2008}
D.~J.~Love, R.~W.~Heath, V.~K.~N.~Lau, D.~Gesbert, B.~D.~Rao, and M.~Andrews, 
``An overview of limited feedback in wireless communication systems,'' 
\emph{IEEE J. Sel. Areas Commun.}, 
vol.~26, pp.~1341--1365, Oct.~2008.

\bibitem{Tse-book}
D.~Tse and P.~Viswanath, 
\emph{Fundamentals of Wireless Communication}. 
Cambridge, U.K.: Cambridge Univ. Press, 2005.

\bibitem{Beck}
A.~Beck, 
\emph{Introduction to Nonlinear Optimization: Theory, Algorithms, and Applications}. 
SIAM, 2014.

\bibitem{Sharif-2005}
M.~Sharif and B.~Hassibi, 
``On the capacity of MIMO broadcast channels with partial side information,'' 
\IeeeTIT, vol.~51, pp.~506--522, Feb.~2005.

\bibitem{Hur-2013}
S.~Hur \emph{et al.}, 
``Millimeter wave beamforming for wireless backhaul and access in small cell networks,'' 
\IeeeTCOMM, vol.~61, pp.~4391--4403, Oct.~2013.

\bibitem{Schwarzbeck.2026}
J.~Schwarzbeck, R.~Neuder, M.~Späth, and A.~Jiménez-Sáez,
``Scalable mm-wave liquid crystal reconfigurable intelligent surfaces based on the delay line architecture,'' 
\emph{arXiv:2601.11307}, 2026.

\bibitem{Jain.1984}
R.~Jain, D.~Chiu, and W.~Hawe, 
``A quantitative measure of fairness and discrimination for resource allocation in shared computer systems,'' 
\emph{DEC Research Report TR-301}, Sep.~1984.

\end{thebibliography}
\end{document}